\documentclass{aastex}
\usepackage{emulateapj5}
\usepackage{lscape}
\usepackage{onecolfloat}   

\newcounter{affil}

\newcommand{\logg}{\ensuremath{\log g}}

\newcommand{\msun}{\ensuremath{\rm M_\odot}}

\newcommand{\beq}{\begin{equation}}
\newcommand{\eeq}{\end{equation}}
\newcommand{\beqa}{\begin{eqnarray}}
\newcommand{\eeqa}{\end{eqnarray}}

\newcommand{\tableskip}{\\[-6pt]}

\begin{document}
\twocolumn[

\submitted{Astrophysical Journal Supplements, in press}
\lefthead{SDSS White Dwarf Catalog}

\title{A Catalog of Spectroscopically Confirmed White Dwarfs 
from the Sloan Digital Sky Survey Data Release 4}
\author{
Daniel J.\ Eisenstein\altaffilmark{1,9},
James Liebert\altaffilmark{1},
Hugh C.\ Harris\altaffilmark{2},
S.J.\ Kleinmann\altaffilmark{3,4},
Atsuko Nitta\altaffilmark{3,4},
Nicole Silvestri\altaffilmark{5},
Scott A.\ Anderson\altaffilmark{5},
J.C.\ Barentine\altaffilmark{3},
Howard J.\ Brewington\altaffilmark{3},
J.\ Brinkmann\altaffilmark{3},
Michael Harvanek\altaffilmark{3},
Jurek Krzesi\'nski\altaffilmark{3,6},
Eric H.\ Neilsen, Jr.\altaffilmark{7},
Dan Long\altaffilmark{3},
Donald P.\ Schneider\altaffilmark{8}, 
Stephanie A.\ Snedden\altaffilmark{3}
}

\begin{abstract}
We present a catalog of 9316 spectroscopically confirmed white dwarfs
from the Sloan Digital Sky Survey Data Release 4.  We have selected
the stars through photometric cuts and spectroscopic modeling, backed
up by a set of visual inspections.  Roughly 6000 of the stars are new
discoveries, roughly doubling the number of spectroscopically
confirmed white dwarfs.  We analyze the stars by performing temperature
and surface gravity fits to grids of pure hydrogen and helium atmospheres.
Among the rare outliers are a set of presumed helium-core DA white dwarfs with
estimated masses below $0.3\msun$, including two candidates that may be
the lowest masses yet found.  We also present a list of 928 hot subdwarfs.
\end{abstract}

]

\altaffiltext{1}{Steward Observatory, University of Arizona,
		933 N. Cherry Ave., Tucson, AZ 85121}
\altaffiltext{2}{United States Naval Observatory, Flagstaff Station,
                P.O.\ Box 1149, Flagstaff, AZ 86002}
\altaffiltext{3}{Apache Point Observatory,
                P.O.\ Box 59, Sunspot, NM 88349}
\altaffiltext{4}{Subaru Telescope, 650 N. A'Ohoku Place, Hilo HI, 96720}
\altaffiltext{5}{University of Washington, Astronomy Department,
                Seattle, WA}
\altaffiltext{6}{Mt. Suhora Observatory, Cracow Pedagogical University, ul.\ Podchorazych 2,
	    30-084 Cracow, Poland}
\altaffiltext{7}{Fermilab National Accelerator Laboratory,
     P.O. Box 500, Batavia, IL 60510}
\altaffiltext{8}{Department of Astronomy and Astrophysics,
                Pennsylvania State University, University Park, PA 16802}
\altaffiltext{9}{Alfred P.~Sloan Fellow}

\section{Introduction}
The Sloan Digital Sky Survey (SDSS) is greatly expanding the number
of spectroscopically confirmed white dwarf stars.
\citet{Har03} presented a catalog of 269 white dwarfs from
the Early Data Release, and \citet{Kle04} followed this
with a catalog of 2551 white dwarfs from SDSS Data Release 1.
More specialized SDSS catalogs have been presented for 
magnetic white dwarfs \citep{Sch03,Van05}, 
white dwarf plus M star binary systems \citep{Ray03,Smo04,Sil05}, 
and cataclysmic variables \citep{Szk02,Szk03,Szk04,Szk05}.  
The size of the SDSS white dwarf catalog has 
significantly expanded samples of rare white dwarfs, including 
ultra-cool white dwarfs \citep{Har01,Gat04,Kil05},
stars with atomic carbon and oxygen \citep{Har03,Lie03}, 
pulsating DAV stars \citep{Muk04,Mul05}
and DBV stars \citep{Nit05},
hot DO stars \citep{Krz04},
very low accretion rate magnetic cataclysmic variables
\citep{Szk03b},
low-mass degenerate helium core stars \citep{Lie04}, 
and hot DB stars in the ``DB gap'' \citep{Eis05db}.  

Here we present a catalog of white dwarfs and hot subdwarf stars
from the SDSS Data Release 4 \citep{Aba05},
which presents 800,000 spectra
from 4783 square degrees.  
We use automated techniques supplemented
by visual classification to select 13,000 candidates.  
An extensive analysis of these objects yields
9316 white dwarfs (WD), including 8000 DAs, 713 DBs, 41 DOs or
PG1159 stars,
289 DCs, 104 DQs, and 133 DZs.  We find 928 hot subdwarf stars (SD),
building on the samples from \citet{Har03} and \citet{Kle04}.
We fit the WDs to grids of model atmospheres and 
present temperatures and surface gravities of the DA and DB stars.
We also present 774 duplicate spectra of WDs and 60 duplicate spectra
of SDs.

This catalog is not complete, even within the SDSS spectroscopic
sample, as we have intentionally focused on the ``easy'' white dwarfs,
namely those that fall in the blue color region of DA and DB stars.
Low temperature stars that fall too near the colors of A and F
stars have not been searched.  Cataclysmic variables have been
excluded, as they have been separately cataloged.  White dwarf plus
M star binaries are included where easily identified, but the catalog of 
\citet{Sil05} is more complete and better analyzed.  Magnetic 
WDs are sufficiently heterogeneous to fool our automated techniques,
but we include them where possible.  DC stars are difficult to
positively identify, although we do find a fair number.

Although completeness was not our goal, we expect that the catalog
does include nearly all of the SDSS white dwarf spectra for the common classes 
of hotter DA, DB, and DO.
We have attempted to find DQ and DZ stars, although this effort is often limited by
the signal-to-noise ratio of the spectra.

\section{The SDSS}
\label{sec:sdss}

The SDSS \citep{Yor00,Sto02,Aba03,Aba04,Fin04,Aba05a,Aba05}
is surveying $10^4$ square degrees of high-latitude sky
in 5 bandpasses: $u$, $g$, $r$, $i$, and $z$ \citep{Fuk96,Gun98,Gun05}.
The images are processed \citep{Lup01,Sto02,Pie02} and calibrated
\citep{Hog01,Smi02,Tuc05} to produce 5-band catalogs, from which 
galaxies \citep{Eis01,Str02}, quasars \citep{Ric02}, 
and stars are selected for follow-up spectroscopy.
Spectra covering 3800\AA\ to 9200\AA\ with resolution of 1800
are obtained with twin fiber-fed double-spectrographs. 

The SDSS has obtained many spectra of white dwarfs and other blue
stars \citep{Har03,Kle04} due to a variety of target selection
catagories.
See \citet{Har03} and \citet{Kle04} for further discussion,
plots of white dwarfs in SDSS color space, and representative
spectra.  We will discuss the completeness of the spectroscopic
targeting in \S~\ref{sec:complete}.

In addition to the standard survey spectroscopy, DR4 includes 
spectra from the special programs that have been executed over
the course of the survey.  Most of these focus on the equatorial
stripe in the South Galactic Cap.  We have included these spectra
in our search as well.  See \citet{Aba05} for descriptions of the 
programs.  Some programs, being focused on blue stars or quasars, 
are reasonable hunting grounds; others have negligible yield.  
The special programs are indicated by setting the leading bit
of the targeting flags.

The SDSS photometric zeropoints are close to the AB convention \citep{Fuk96},
but do not reproduce it exactly.  We quote all of our photometry
on the SDSS system rather than the AB system.  However, when fitting models,
we do include the following AB corrections:
$u_{\rm AB} = u_{\rm SDSS}-0.040$, $i_{\rm AB} = i_{\rm SDSS}+0.015$,
$z_{\rm AB} = z_{\rm SDSS}+0.030$, with no correction in $g$ and $r$.
These corrections were derived from comparison to the HST STIS
spectrophotometric calibration stars \citep{Boh01} as well as from
analysis of the colors of DA white dwarfs in the SDSS.  We note that
of these corrections, only the $u$-band one matters for any of the WD
fitting.  The reason for the $u$ band discrepancy has been identified: the
definition of the zeropoint in \citep{Fuk96} did not include the effects
of atmospheric absorption on the shape of the response function.
The AB corrections remain uncertain at least at the 1\% level; we
are continuing to work on this issue.

\section{Catalog Selection and Inspection}
\subsection{Initial Selection}
\label{sec:select}

White dwarfs hotter than roughly 8000 K are distinct in SDSS colors
from the much larger samples of galaxies, quasars, and F stars that
the SDSS targets for spectroscopy.  The primary exception are the 
white dwarfs with late-type companions, which still remain distinct
from the stellar locus in color, albeit in a different region.

\begin{figure}[tb]
\plotone{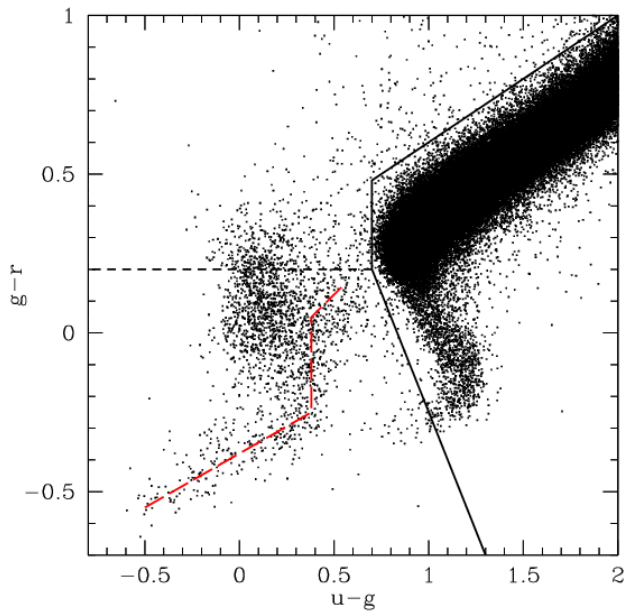}
\caption{\label{fig:ugrall}%
The $u-g$ and $g-r$ colors, dereddened, of blue point sources
with clean photometry from about 200 square degrees of the SDSS.
The black solid and short-dashed lines show our selection regions,
described in \S~\protect\ref{sec:select}.  The red long-dashed line
shows the approximate location of the DA color locus.  The region
of dense points is the main sequence, extending blueward in $g-r$
to the upper main sequence and blue horizontal branch, before turning
blueward in $u-g$ and extending (at much lower densities) toward
the hot subdwarfs (shown in Figure \protect\ref{fig:ugr}).
The cloud of objects at blue $U-g$ and $g-r\sim0.2$ are mostly quasars.
}
\end{figure}

Therefore, we begin with a color selection of all objects with 
SDSS spectroscopy that have $u<21.5$ and colors in one of the
following two regions:
\begin{eqnarray}
&& -2<u-g<0.833-0.667(g-r), \\
&& -2<g-r<0.2 
\end{eqnarray}
or 
\begin{eqnarray}
&& 0.2<g-r<1, \\
&& |(r-i)-0.363(g-r)|>0.1, \\
&& u-g<0.7 {\rm\ or\ } u-g<2.5(g-r)-0.5.
\end{eqnarray}
The first box selects all of the stars that are bluer in $g-r$ than the 
quasars and bluer in $u-g$ than the main sequence A stars.
The second box runs blueward of the stellar locus in $u-g$ at moderate
$g-r$, in order to find WD+M stars.  Note that we require that the
color not fall on the stellar locus in $g-r$ and $r-i$; this 
will result in some incompleteness.  Figure \ref{fig:ugrall}
shows the selection region in $u-g$ and $g-r$ colors relative
to the main stellar locus.
All of these magnitudes are PSF photometry, dereddened in full
from the \citet{SFD} extinction map.  This may overcorrect stars
that are in front of some of the Galactic dust, but this has
the preferred tendency to bring nearby objects into the sample.
We excluded all regions with $r$-band extinction above 0.6
($A_g>0.827$ mag);
this removes only a few low-latitude plates outside of the survey region.

We further require that the object pass at least one of the 
following three cuts based on
the two SDSS redshift pipelines (specBS and spec1d).  1) For specBS, we require
$|z|<0.003$ and that the classification not be that of a galaxy.
2) For spec1d, we require that zwarning bit 1 not be set and additionally
that no band of photometry be flagged as {\tt Interp\_Center}, {\tt CR}, 
{\tt Edge}, or {\tt Satur} \citep{Sto02}.
3) Alternatively, objects could enter the sample from spec1d if their
redshift was greater than 0.003, the photometry flags were not set,
and their proper motion from USNO-A was greater than $0.3''$ per year. 
The photometry flags were not checked in option 1, nor was the object required
to be a point source.
Unfortunately, due to data availability at the time,
options 2 and 3 were only queried for point sources in Data Release 3.
Option 1 supplied 12904 candidates.
Options 2 and 3 supplied another 978 new candidates, for a total
of 13882.
When restricted to the DR4 region, we analyzed 13641 candidates.
1028 of these were duplicates, which we excluded from the primary analysis.

This selection will not capture all of the white dwarfs in the 
SDSS spectroscopic catalog, but instead focuses on the regions
that include most of the spectroscopic WDs.
The obvious failure modes are as follows.
1) White dwarfs cooler than about 8000 K are sufficiently close to the 
stellar locus that we have not attempted to dig them out.
The SDSS photometry is of sufficient quality to cut closer than
we have here, but our automated spectral analyses suffer enough at
low temperatures that we would have to check many thousand candidates
by eye.  
2) White dwarfs in binaries with brighter companions can have colors
more typical of the stellar locus and hence avoid our selections.
3) Rare classes of white dwarfs may have been given a non-zero
redshift by the spectroscopic pipelines if they do not match any
of the spectral templates.  For example, magnetic WDs can often 
fool the pipelines.  We actually do recover many magnetic WDs,
but we are likely not complete.  DC white dwarfs also can be
dropped from the catalog if the pipelines interpreted a weak
noise feature as a low-confidence extragalactic redshift.
4) DC white dwarfs are difficult to confirm in any case.

Our view is that the cooler WDs, being spectroscopically 
more complicated, are better suited to proper motion analyses.
These are being pursued by \citet{Har05} and \citet{Kil05}.
Similarly, the WD+M pairs are being cataloged by \citet{Sil05}.

\subsection{Spectroscopic Analysis}
Thus far, we have only selected blue stars, not white dwarfs.
We next fit the 13641 candidates to a grid of white dwarf atmospheres
using the autofit method described in full in \citet{Kle04}.
This separates white dwarfs from lower gravity stars.

In brief, the autofit method performs a $\chi^2$ minimization
of the spectrum and photometry of each candidate relative to a 
model atmosphere grid, thereby yielding a temperature and
surface gravity estimate as well as a classification of 
helium or hydrogen atmosphere.  We use two sets of atmospheres
provided by D.~Koester \citep{Fin97},
one pure hydrogen, the other pure helium.  Both are computed
assuming local thermodynamic equilibrium (LTE) populations.
The hydrogen atmospheres
run from 6000 K to 100,000 K and $\log g = 5.0$ to 9.0.
The helium atmospheres run from 10,000 to 40,000 K and $\log g=7.0$
to 9.0.  All model spectra are convolved to the spectral resolution
of the SDSS data.  

The $\chi^2$ statistic includes
both spectral and photometric residuals.
We include the spectrum from 3900\AA\ to 6800\AA.  
The spectral continuum is controlled by marginalizing over an unknown
polynomial of 5th order that multiplies the model spectrum.
We run autofit on the observed, not dereddened spectra, but
this order of polynomial is sufficient to be completely 
degenerate with the uncertain reddening or with any residual
effect of atmospheric dispersion at the fiber entrance.
The photometric fit marginalizes over the distance to the star
and over extinction values ranging from zero to the \citet{SFD}
prediction, with penalities increasing thereafter.
We apply the AB zeropoint corrections described in \S~\ref{sec:sdss}.
Masked pixels in the spectrum are excluded,
as are any photometric bands with suspicious warning flags.
The star is designated as a hydrogen or helium star based
on the best-fit $\chi^2$ in the two grids.
Errors on the fitted parameters are computed by taking moments
of the likelihood function $\exp(-\chi^2/2)$ within the
chosen grid.

\begin{figure}[tb]
\plotone{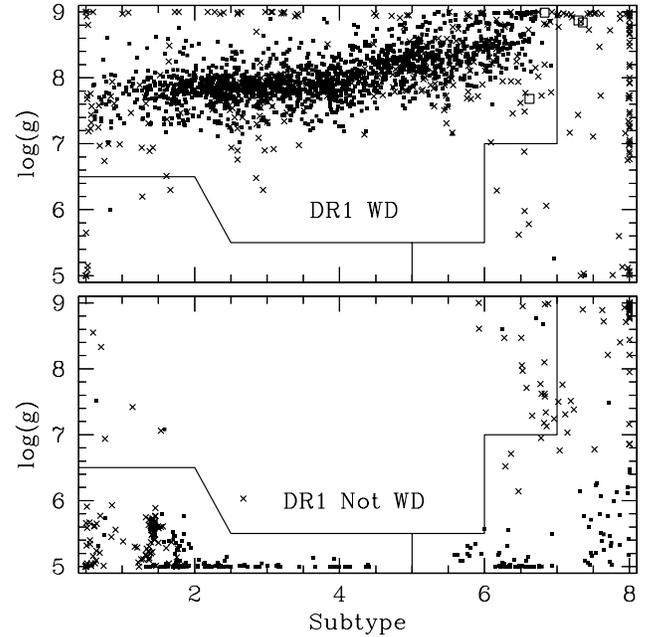}
\caption{\label{fig:DAall}%
The autofit temperature and gravities for all of the stars in our parent list
that were fit to hydrogen atmospheres and that were 
visually classified in the DR1 WD catalog \protect\citep{Kle04}.  
The upper panel shows those classified as
white dwarfs; the lower panel shows those classified as non-WD.  In each panel, lines divide the 
plot into three regions marking the autofit
first-pass classification: upper-left for DA white dwarfs, lower-left for 
subdwarfs, right for unknown.  The crosses show objects selected for visual
inspection.  The squares show objects for which the autofit result would 
have been accepted were we not applying the visual classifications from 
\protect\citet{Kle04}.  The figures show that nearly all of the 
automatically classified white dwarfs (squares in the upper left) are
white dwarfs: there are only 5 squares in the upper left of the bottom
panel.  Moreover, nearly all of these stars were visually classified as DAs:
in the top panel, we mark DAs with solid squares and other WDs with 
open squares.  The upper left region of the top panel contains about 1700
autoclassified stars, all but 2 of which are DAs.  Non-DA white dwarfs
may fall in that region, but they were selected for inspection.
Squares in the other regions in the top panel mark white dwarfs that were
not recovered, having been accepted as other types.  These are also rare.
}
\end{figure}

\begin{figure}[tb]
\plotone{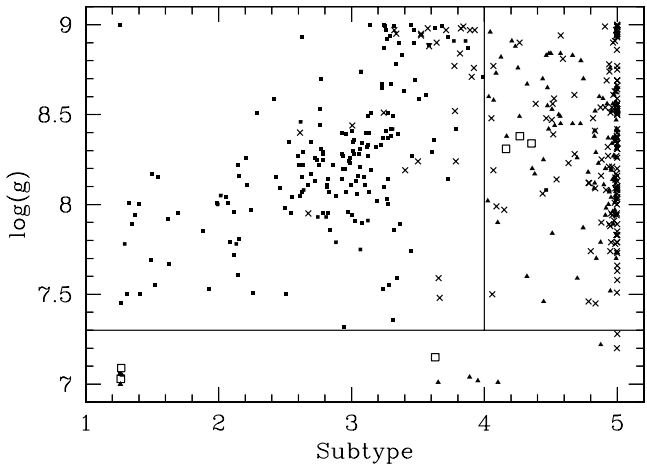}
\caption{\label{fig:DBall}%
The autofit temperature and gravities for all of the stars in our parent list
that were fit to helium atmospheres and that were visually classified in the 
DR1 WD catalog \protect\citep{Kle04}.  
The lines show the autofit first-pass classifications: upper-left for DB
white dwarfs, bottom for subdwarfs, upper-right for unknown.
The solid squares are visually classified DBs that were selected for inspection
(we selected the whole first-pass DB region for inspection so as to improve
the catalog for the rarer DBs).  The open squares are visually classified DBs
that were not selected for inspection and hence would have been missed.
The crosses are other objects selected for inspection; the triangles are
other objects rejected by autofit.  The other objects are often 
other classes of white dwarfs or cataclysmic variables, since low-temperature
helium atmospheres have featureless continua that become the best-fit
model in our limited fitting space.
}
\end{figure}

We find that the autofit temperatures and surface gravities 
do a credible job, even unsupervised, of distinguishing between 
white dwarfs and other types of stars.
This is shown in Figures \ref{fig:DAall} and \ref{fig:DBall},
which show the distribution of temperatures and gravities 
of the hydrogen and helium stars that were in the DR1 WD
catalog \citep{Kle04}.  The visual classification from 
\citet{Kle04} is shown in the two panels of Figure \ref{fig:DAall}
and in the point type in Figure \ref{fig:DBall}.  The separation
of the white dwarf locus is clear.  Subdwarfs tend to fall
at low gravity and moderate temperatures; F and A stars
fall to the low temperature extreme.  Other weak line stars
and CVs fall at low temperatures, particularly in the helium
grid, where the model lines are very weak.  Notably, DZ, DC,
and DQ stars often migrate to low fitted temperatures; the next
section will describe how we attempt to recover these stars.

We therefore designate regions in the temperature-gravity space
to indicate our first-pass classification.  These regions are
drawn in Figures \ref{fig:DAall} and \ref{fig:DBall}.
The high-temperature, high-gravity region are called DA and DB,
respectively; the high-temperature/low-gravity region are 
hot subdwarf candidates; and the low-temperature region are
left as unknown.

\subsection{Inspections}

We then subject these classifications to a large series of additional
tests.  Failing these tests leads to visual examination, which can
reject the object as a white dwarf, alter its classification, or rescue
it back from the unknown pile.

Several of these tests involve searching for anomalous spectral
features.  We do this by defining narrow windows at the relevant
wavelength and measuring the equivalent width of the residuals
of the spectrum relative to the best-fit model (i.e. we use the
model to define the continuum rather than interpolating between
from neighboring windows).  We also estimate the statistical
error in the equivalent width using the quoted spectral errors.
This ignores the contribution of fluxing errors to the scatter,
but this approach is conservative.
In this manner, we search for 
HeI 4471, HeII 4686, $H\alpha$, $H\beta$, CaII 3931, Mg b, 
and the $C_2$ Swan band at 5175 \AA.  

For the first-pass DA stars, we flag any object whose
$\chi^2$ from the photometry alone exceeds 50 (on 3 degrees
of freedom).  These are often DA+M stars, where the $i$ and
$z$ photometry is far redder than the blue WD fit.
We then search for HeI 4471, HeII 4686, and Swan in emission or
absorption and for Mg b or CaII 3931 in absorption; these can
find DZ, DQ, DAB, DAO, and CV stars as well as subdwarfs.
We flag any star with $\log g<7$ or $T>84,000$ K, as well as 
$\log g>8.9$ and $T>10,000$ K, as these outliers are suspicious.  
We flag stars that had reduced $\chi^2$ greater than 1.5,
or signal-to-noise ratio (SNR) less than 2 per spectroscopic pixel, 
or 2 or fewer bands of good photometry.  We also flag
stars with SNR above 4 whose $\chi^2$ in the helium and
hydrogen fits differ by less than 5\%. 

The approach is the same
for the first-pass DB stars, but we search
for $H\alpha$ and $H\beta$ in absorption instead of HeI 4471.  
After finishing these higher priority inspections, we continued
and inspected all of the first-pass DB stars.

For the first-pass hydrogen subdwarfs, we flag stars with
gravities above 6.0, as these might be DO stars.  
For the first-pass helium subdwarfs,
we flag those that lack $H\alpha$ and $H\beta$ detections,
as these might be misclassified DB stars.  We further
inspected all of the subdwarfs with $u-g<0.2$, again
to look for DO stars.

For the first-pass hydrogen unknowns, our goal is to recover
DA+M stars.  There are large numbers of F and A star 
contaminants that tend to pin against the grid extremes in
temperature and gravity.  We inspect stars that lie away from the  
edge of the grid.  We also flag stars if they have a jump
in their spectrum at 8100 \AA, as this might signal an M star.

The first-pass helium unknowns are an interesting set because
the lowest temperatures in the helium grid have no absorption lines,
so any stars without H or He I absorption end up here.
We flag stars with $H\alpha$ and $H\beta$ in emission; these
are often CVs.  We flag stars with Ca II in absorption to find
DZs and Swan bands to find DQs.  

This selects about 3000 stars for visual inspection, some of which
were already inspected in \citet{Kle04}.  The visual classification
of these stars is given in the catalog.  However, given the size
and heterogeneity of the sample, it is inevitable that there will
be errors and uncertainty in the visual classifications too,
particularly among certain difficult boundaries.

We expect that the stars that are not visually inspected 
are a reasonably pure sample, but inevitably some mistakes
will have crept in.  Figures \ref{fig:DAall} and \ref{fig:DBall}
show the stars not selected for inspection that overlap with
the \citet{Kle04} sample.  The DA and DB loci are very pure,
and we expect that we have found most of the isolated DA, DB,
and DZ stars above 10,000 K in the SDSS spectroscopic sample.  
Our selection has
not been tuned to WD+M binaries and we have lost these stars both
in the photometric selection and the spectroscopic classification.

We find that the searching for anomalous emission and absorption
has produced sizable lists of DZ, DQ, DAB, DAO, CV, and magnetic WD stars,
but we have not made any assessment of the completeness of
these samples.  

Our catalog of hot (sdB/sdO) subdwarfs is likely reasonably complete,
but we have not inspected it to remove interlopers.  The
temperatures and gravities from autofit surely correlate with
the true values, but they are inaccurate because
we have not used mixed hydrogen/helium atmospheres.  
We present the fitted parameters only as a secondary aid to sorting through the catalog.  
Importantly, our dividing line between the subdwarf list
and the unknown list was not drawn to match a robust temperature
scale nor to match any particular visual classification standard.
Hence, the subdwarf sample should not be used for statistical work
without further analysis of the temperature scale and completeness.

At temperatures below 10,000 degrees, the Balmer lines in DA white dwarfs
become rather narrow, and it becomes difficult to distinguish between 
such stars and cooler subdwarfs and blue horizontal branch stars. 
Normally, the strength of the higher Balmer lines would be used as
a surface gravity diagnostic, but this is not always possible because
the SDSS spectra become noisy toward the UV.  There are likely 
some DA stars that have been misclassified as subdwarfs and horizontal
branch stars, and vice versa.

Similarly, distinguishing between helium-dominated subdwarfs and
DB white dwarfs can be subtle, particularly in low signal-to-noise
ratio spectra.  One can expect a small level of misclassification.
However, we have visually inspected all of the DB stars, and are
confident that any contamination is small.

\section{Classification Schemes}

Our visual classifications are based on the systems presented in 
\citet{Har03} and \citet{Kle04}.
We classify white dwarfs in the standard way as DA, DB, DC, DH, DO,
DQ, and DZ.  If secondary lines are present, we attach a second
letter.  For example, a DAB would have dominant hydrogen lines and
weaker neutral helium lines.

In all cases, a colon indicates uncertainty in the classification.  
We use a space to indicate when the uncertainty should be applied
only to a portion of the classification.
For example, ``DA H:'' would mean a DA star
with possible magnetic contributions, while ``DAH:'' means that even
the DA portion is not firm.  We note in particular that by
``DBA:'' and ``DAB:'' we are expressing uncertainty
with the interpretation as a white dwarf (as opposed, usually, to
a hot subdwarf).  We write ``DB A:'' and ``DA B:'' if we mean that
only the presence of the secondary lines is uncertain.

We write ``DQhot'' to indicate that the carbon lines are atomic
rather than molecular \citep{Har03,Lie03}.  Some of these stars, 
e.g., SDSS J220029.08-074121.5, are hot enough to show once-ionized
carbon, similar to some of the stars in \citet{Lie03}.  

We write ``PG1159'' to indicate a hot star that is showing highly
ionized lines of carbon and oxygen.  We have included these as 
white dwarfs, although these stars are known to span a range of
gravities and the evolutionary tracks are continuous between the 
hottest subdwarfs and hot white dwarfs.  

DC stars are generically hard to classify securely.  In cases where
SDSS has a reasonably high signal-to-noise ratio featureless spectrum
with no radio or X-ray counterpart from the FIRST and ROSAT surveys
and the object has a statistically significant proper motion, then we assign the
star a DC classification.  However, the SDSS has many examples of stars with
noisier spectra where weak features might be missed.  If a star had
a significant proper motion and lacked a radio or X-ray counterpart and if the 
spectrum looked encouragingly but not convincingly featureless, we 
assigned a DC: classification.  The DC: class is likely more permissive
than what we would require of a DA: or DB:, but we didn't want to lose
these stars from WD consideration.  Further, at low signal-to-noise 
ratio, the recovery of white dwarfs is surely biased away from DCs and
toward white dwarfs with strong features.  This bias should be 
recognized in statistical studies.  Likely it can only be avoided by
using a reasonably high signal-to-noise ratio or magnitude cut, so 
as to try to give definitive identifications to nearly all stars.

In this catalog, we have written nearly all companion stars as simply
``+M''.  Although the companion is almost certainly a dwarf, we have 
not written ``dM''.  We have not rigorously typed the star to be a M 
star.  Indeed, sometimes the evidence for the companion is merely 
emission; we have not indicated such.  A colon indicates that the 
evidence for the companion is not secure.  We refer the reader to the
paper by \citet{Sil05} for a complete treatment of WD+M binaries, 
including finer classification,
joint parameter fits, and detailed searches for emission.
The \citet{Sil05} classifications are to be preferred to our simple ``+M''
scheme.

For the hot subdwarfs, we have sorted the stars into three classes:
SDO, SDB, and Helium-rich SDB (``HeSDB'').  However, we caution
that our classification has not been particularly systematic.  The
SDSS hot subdwarfs would benefit from a more quantitative classification
scheme, e.g., following \citet{Jef97}.

For some of the hot subdwarfs we write ``+MS'' to indicate a
main sequence companion; hot subdwarfs are luminous enough to hold their own 
with a F or G star, rather than merely a late-type star.  However,
this classification is very incomplete.  As discussed below, the $u-g$ vs. $g-r$
diagram reveals a clear second locus of classified subdwarfs 
that is clearly due to companions, but we have not been systematic 
in applying this information.

The automatic classifications are limited to the typing: DA, DB, SD,
or other.  We add the word ``auto'' to the classification if a human
has not confirmed the classification.  There is no binary statement of
uncertainty; rather, we refer the user to the temperature and surface
gravity values and errors.  Obviously, large errors can indicate
uncertainty not just in the parameters but in the basic interpretation.
This is particularly true in the case of surface gravity.  The errors on
the temperature are sometimes quite large because the fit is straddling
the instability strip and has two local minima (i.e., DA3 and DA5).
This does not call the DA interpretation into doubt, but merely indicates
that the temperature is uncertain.

Finally, it should be noted that SDSS spectroscopy extends to rather
low flux levels ($g\approx20.5$) and the spectral signal-to-noise 
ratio of the faintest objects is too poor (as low as 3 per spectroscopic pixel)
to permit robust classification.  We have attempted to indicate
this uncertainty with colons and with the quoted autofit errors.
But clearly there will be a strong magnitude and signal-to-noise ratio
bias in the recovery of secondary features.

\section{Catalogs}
\subsection{General Properties}

We present the catalogs as two tables, one for white dwarfs, the other
for hot subdwarfs.  The tables include the classifications, the 
identification information for both imaging and spectroscopy, 
the photometry, astrometry, and autofit information.  The catalog
format is given in Table \ref{tab:cat} but the lines are too long
to give in-print samples.  

A listing of the classifications, along with the number of stars in
each, is given in Tables \ref{tab:wdsuper}, \ref{tab:wd}, and \ref{tab:sd}.
Table \ref{tab:wdsuper} is simply a summary of the dominant classification,
meaning the first two letters ignoring any secondary information.
The full diversity of classifications are in \ref{tab:wd} and \ref{tab:sd}.
Although mixed typed stars are rare, the size of the DR4 catalog does yield
a significant number of them, with many different permutations of the types.
For example, there are 16 of the interesting DZA and DAZ stars 
\citep{Dup92,Zuc98,Zuc03,Koe05}, with 5 uncertain cases and 11 DBAZ/DBZA/DZBA stars.

Figure \ref{fig:ugr} shows the $u-g$ versus $g-r$ distribution
of the stars in the different dominant classifications.  As expected,
the DA stars have a bent locus due to the strong Balmer break at
intermediate temperatures, while the helium-dominated atmospheres
follow a more linear locus close to that of a blackbody.  The WD+M
stars scatter redward in $g-r$ from the single-star locus.  It is
worth noting that the single star loci end at the dashed line
at $g-r=0.2$ rather than the solid line that indicates the boundary
of our selection region.  This is because at $g-r>0.2$ our selection requires 
that the star be off of the main sequence color locus in $g-r$ vs.\ $r-i$.
This cut is satisfied by the binary stars but not the single stars. 
We will postpone discussion of the subdwarf panel until \S~\ref{sec:sd}.

\begin{figure}[tb]
\plotone{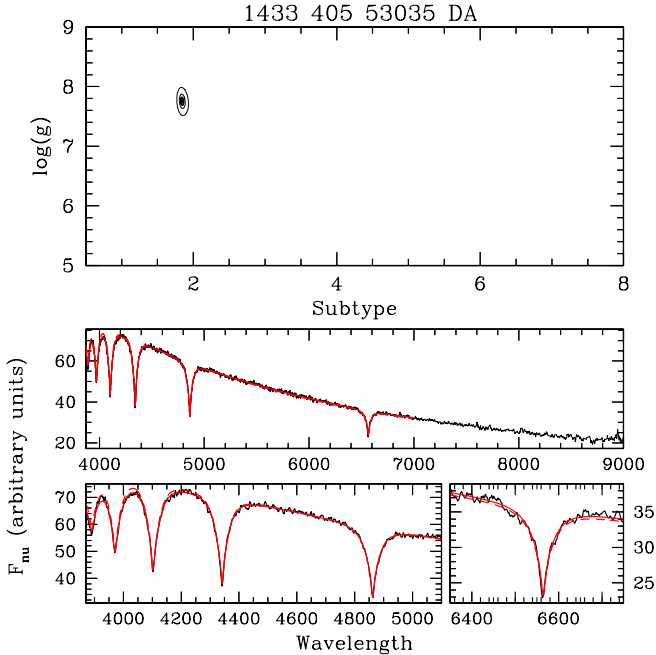}
\caption{\label{fig:autofitDA}%
The spectrum and autofit model for the DA star SDSS J104419.00+405553.0 (plate
1433, fiber 405, MJD 53035).  This star is unusually bright ($g=16.83$)
and high S/N (38 per spectroscopic pixel in the $g$ band). 
The middle panel shows the full spectrum; the lower panels show expanded views
of the Balmer lines.  The data is in black.  The dashed red line is the
best-fit model.  The solid red line is the best-fit model having been
adjusted by the low-order polynomials that represent reddening and fluxing
errors.  The top panel plot the likelihood contours for the autofit modeling
($\Delta\chi^2$ corresponding to 1, 2, 3, 5, and 10-$\sigma$ for a 
two-dimensional Gaussian).
The DA label at the top is not the classification, but merely the statement
that this fitting is for a hydrogen atmosphere.  
}
\end{figure}

\begin{figure}[tb]
\plotone{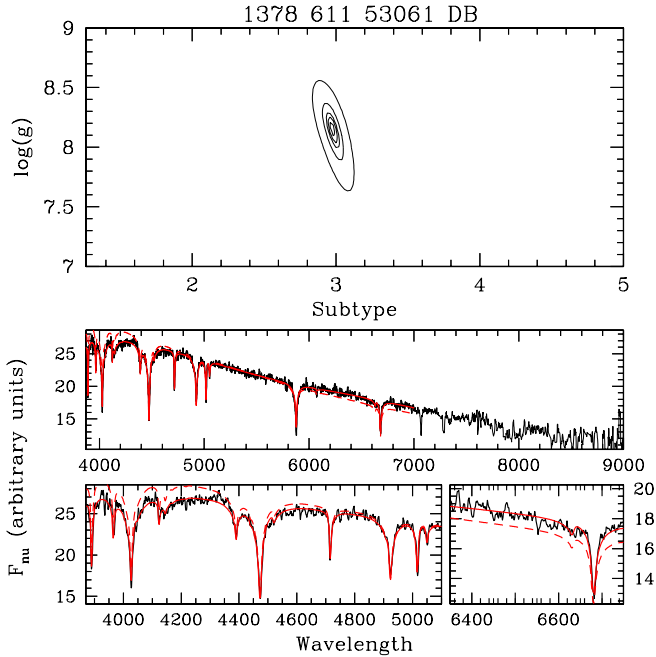}
\caption{\label{fig:autofitDB}%
The spectrum and autofit model for the DB star SDSS SDSS J140227.19+403922.2 (plate
1378, fiber 611, MJD 53061).  This star is unusually bright ($g=17.78$)
and the spectrum has a high S/N ratio (29 per spectroscopic pixel in the $g$ band). 
The lines and panels are the same as Figure \protect\ref{fig:autofitDA}.
The DB label at the top is not the classification, but merely the statement
that this fitting is for a helium atmosphere.  
}
\end{figure}

The spectra themselves are available from the SDSS Data Release web 
site\footnote{http://www.sdss.org/dr4/}.
Each autofit model generates a summary figure, showing the spectrum,
the best fit, and the likelihood contours in temperature and surface
gravity.  These figures are available 
on-line\footnote{http://das.sdss.org/wdcat/dr4/};
examples of
a bright DA and bright DB are shown as Figures \ref{fig:autofitDA} and 
\ref{fig:autofitDB}.  The best-fit model is shown in the red dashed 
line.  The red solid line shows the model modified by a multiplicative
5$^{\rm th}$-order polynomial.  Typically some modification is appropriate
to account for extinction or spectrophotometry residuals.  The spectra
themselves are shown as observed, not dereddened, so it is quite common
for the spectrum to be somewhat redder than unextincted model.
However, if the broadband model is very different than the data, or 
the modified model is a poor fit to the data, then this is of course
a warning that the autofit parameters are not trustworthy.

In table \ref{tab:dr1orphans}, we list the stars from the SDSS DR1 white dwarf
catalog of \citet{Kle04} that did not make the list here.  There
are 73 such stars, about 3\% of the \citet{Kle04} sample.  These stars
are missing not because we doubt their classification but because they
did not pass our initial selection cut.  In a few cases, this is because
the extinction was large or because the photometry was bad.  Some of the
low temperature stars or binary systems simply didn't pass our color cuts.
The primary reason, however, was that the system was not identified as being
at redshift $|z|<0.003$ by either pipeline.  This is common in the case of
DC stars, when the pipeline may interpret a weak noise feature as
a extragalactic redshift, even if it reports it as low confidence.  There
are about a dozen normal WDs that fool the pipeline as well.  For this work,
we did not inspect every extragalactic redshift for a blue point source,
even those at low confidence.  There are many quasars that satisfy those
cuts, and we chose not to devote that much effort to rectify a small incompleteness.
Future catalogs might find reasonable paths around this, particularly by
working with the QSO catalog team \citep{Sch05}.

In 52 cases, DR4 includes a spectrum of a white dwarf without the
relevant photometry.  We have chosen to include these objects in
the catalogs despite the missing information.  In most cases, the
star is part of a special program or a bonus plate located outside
of the nominal survey area; such imaging was not included in DR4.
In the few remaining cases, the star was blended with other nearby
objects in such a way that it was recovered and targeted in a early
version of the photometric reductions but not identified as a
separate object in later reductions and hence missed in the astrometric
matching.

We have matched our catalog to the listing of literature white dwarfs
compiled by \citet{McC99} (updated as of August 2005).  We find 2343 matches.
However, 2179 of these matches were published in the SDSS DR1 white
dwarf catalog \citep{Kle04}, which was then incorporated into the updated
\citet{McC99} list.  Only 162 of the matches are to stars that were not
in \citet{Kle04}.  In total, the DR4 catalog includes 6159 white dwarfs
that were not in \citet{Kle04}, \citet{Sil05}, or the \citet{McC99}
list, although it is possible that a handful of these stars are already
in the literature.  As this number is slightly more than the total in
the \citet{McC99} list, we have slightly more than double the world's
list of spectroscopically confirmed white dwarfs.

In 9 cases matched to the \citet{McC99} list, we classified the spectrum as a subdwarf:
SDSS J074010.50+284120.8 (WD0737+288),
SDSS J082802.03+404008.9 (WD0824+408),
SDSS J095847.23+602147.2 (WD0955+606),
SDSS J102234.91+460058.7 (WD1019+462),
SDSS J122711.17+003328.8 (WD1224+008),
SDSS J152553.46+434127.7 (WD1524+438),
SDSS J153411.09+543345.3 (WD1532+547),
SDSS J154338.68+001202.1 (WD1541+003), and
SDSS J233541.46+000219.4 (WD2333-002).
These 9 have been eye-inspected to confirm the rejection of the white dwarf interpretation.

\subsection{Autofit Results}
\label{sec:autofitres}

We have included the fitted temperatures and gravities for all
stars in the catalog.  However, we stress that these values are only to be used
if the classification is DA or DB (possibly with colons), without
more subtle variations.  If the classification is more complicated,
then this almost surely means that the atmosphere is more complicated
than the simple pure hydrogen and pure helium values used in autofit.
Using the wrong atmosphere grid (e.g., fitting DBA or DZA stars
as a DA) can produce catastrophically wrong results.  Similarly,
in cases where the stars lack Balmer or neutral helium lines, the
method will have fit to the models with the weakest lines, typically
low temperature and high gravities.  This quality assurance decision
is recorded in the catalog by the autofit quality flag.

\begin{figure}[tb]
\plotone{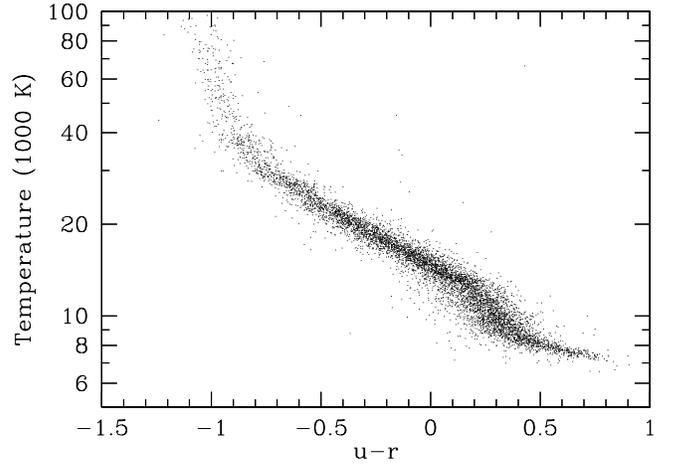}
\caption{\label{fig:da_ur_T}%
The locus of $u-r$ color versus autofit temperature for stars classified
as DA or DA: and fit to a hydrogen atmosphere.
The color is corrected for interstellar extinction assuming
that the star lies behind the \citet{SFD} prediction.  The correlation
between color and temperature is generally excellent, with slightly
more scatter around the instability strip at 11000 K.  While the
photometry is used in the fit, it does not dominate the result;
we regard the agreement as demonstration that the spectroscopic fits
are well-correlated with temperature.  Note that some of the scatter is
simply due to variations in surface gravity.
}
\end{figure}

\begin{figure}[tb]
\plotone{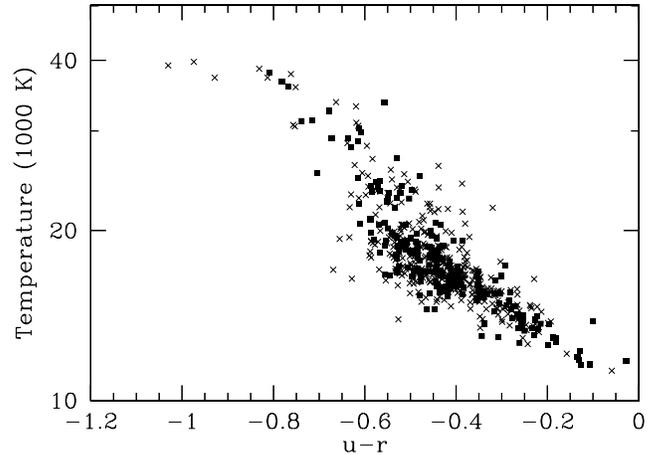}
\caption{\label{fig:db_ur_T}%
The locus of $u-r$ color versus autofit temperature for stars classified
as DB or DB: and fit to a helium atmosphere.
The color is corrected for interstellar extinction assuming
that the star lies behind the \citet{SFD} prediction.  The correlation
between color and temperature is encouraging.  The solid dots are 
spectra with $S/N>10$ per spectroscopic pixel; crosses are those with lower
signal-to-noise ratio.  One sees that there are high S/N cases
across the full temperature range, including stars above 25,000K.
} \end{figure}

\begin{figure}[tb]
\plotone{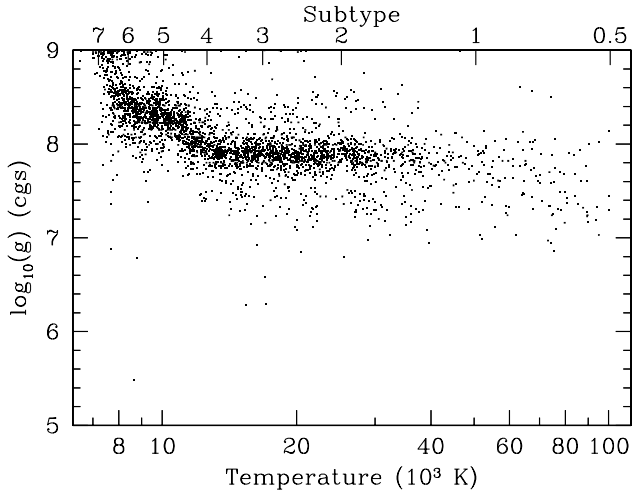}
\caption{\label{fig:da_T_logg_sn10}%
The locus of autofit surface gravity versus autofit temperature 
for stars classified as DA or DA: and fit to a hydrogen atmosphere.
Only stars with $S/N>10$ per spectroscopic pixel in the $g$-band are plotted.
The well-known peak at $\logg = 7.9$ is clear, but the scatter
around this relation is largely real.  One also 
sees the systematic bias toward higher gravities at lower temperatures.
As discussed in the text, this is likely due to systematic errors in
the model atmospheres at low temperatures.  Values very close to 
the boundaries of the model grid (notably at $\logg=9$ or $T=10^5$~K
tend to pull artificially to the boundary and have their error 
underestimated.
}
\end{figure}

\begin{figure}[tb]
\plotone{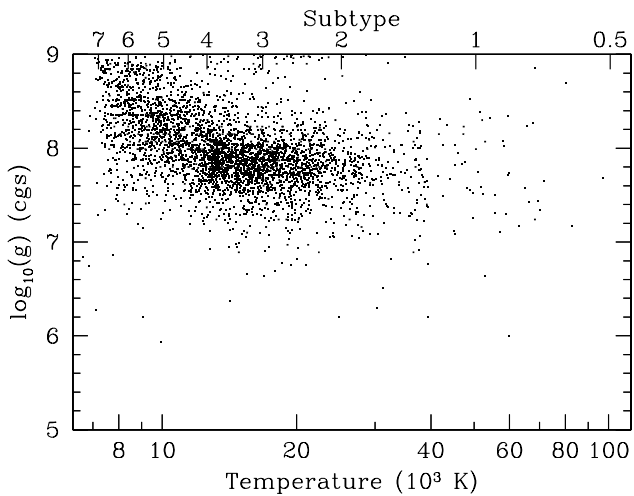}
\caption{\label{fig:da_T_logg_snm10}%
As Figure \protect\ref{fig:da_T_logg_sn10}, but for stars with 
$S/N<10$ per pixel.  The scatter is clearly larger, but the
distribution is otherwise unchanged.
}
\end{figure}

Focusing on the stars with good autofit values, Figures \ref{fig:da_ur_T}
and \ref{fig:db_ur_T} show the autofit temperature versus the $u-r$
color for DA and DB stars.  The correlation is excellent, even at low
spectral signal-to-noise ratio.  Figures \ref{fig:da_T_logg_sn10} and
\ref{fig:da_T_logg_snm10} show the distributions of surface gravity and
temperature for DA stars with signal-to-noise ratios above and below
10 per pixel.  The high signal-to-noise ratio data show a tight locus
reflecting the well-known peak in surface gravity around $\logg=7.9$
\citep{Ber92}.  The lower signal-to-noise data show the same locus albeit
with more scatter.

However, one also sees here the important systematic error that the
surface gravities are overestimated at low temperatures.  Above 10,000
degrees, the peak in gravity falls at the canonical value of $\logg=7.9$.
Below 10,000 degrees the gravities become systematically higher.
This problem was noted and discussed in \citet{Kle04}, but has been
known for decades and is likely a bias in the model atmospheres
rather than the autofit algorithm.
\citet{Ber90} first pointed out that the mean gravities and
masses of DA white dwarfs cooler than the ZZ Ceti instability strip 
($< 11,000$~K) are determined from pure hydrogen atmosphere models
to be higher than for stars above this temperature.
They suggested that the cool atmospheres
could be moderately He-rich, due to incomplete convective mixing with
the underlying He layer.  Helium increases atmospheric pressure and line
broadening in a manner indistinguishable from increased gravity.
However, evolutionary modeling suggests that, if the convection zone of
the H surface layer comes into contact with the helium layer, mixing
should be complete and result in only a trace residual H abundance
\citep{Fon87,Lie87}.
It is also possible that there is a systematic
error in the models, such as in the hydrogen level occupation probability
\citep{Hum88} or in the convection parameterization.  In any case, we
present the autofit results as is, without identifying the source of the
likely bias.

\begin{figure}[tb]
\plotone{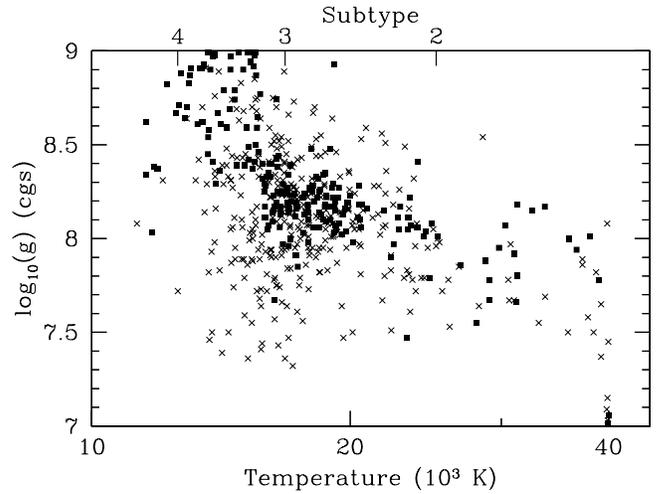}
\caption{\label{fig:db_T_logg_sn10}%
The locus of autofit surface gravity versus autofit temperature 
for stars classified as DB or DB: and fit to a helium atmosphere.
Stars with $S/N>10$ per spectroscopic pixel in the $g$-band are plotted with
solid points; stars with $S/N<10$ are plotted as crosses.
As with the DA stars, the peak in $\logg$ is recovered at higher
temperatures, but lower temperatures have a systematic bias toward higher 
surface gravities.
}
\end{figure}

Figure \ref{fig:db_T_logg_sn10} shows the temperature-gravity locus
for the DB stars.  Again, there is a characteristic peak in the 
surface gravity at higher temperatures, with a rise in gravity 
toward lower temperatures.  This rise was also seen in \citet{Bea96}.
The peak of our surface gravity distribution is somewhat higher than that
of \citet{Bea96}.  Likely this is due to differences in the modeling of
the physics of the helium atmospheres, which is still somewhat unsettled,
particularly in comparison to the hydrogen case.

A systematic problem with the autofit method is that very near the grid
boundaries in temperature and surface gravity, autofit has a systematic
bias to push the fit to the boundary.  In particular, this can cause the
formal error to be very small.  This bias occurs because of the way that the
implementation uses splines to interpolate between the fitting points.
Of course, this is particularly a problem at low temperatures, where
the bias in the model set pushes most of the DA white dwarfs toward
$\logg=9$, where they then pin against the boundary and have incorrectly
small formal errors.

There is an important degeneracy in the fitting of DA and DB stars
because the spectral line strengths reach a maximum near 11,000~K
for DA stars and 25,000~K for DB stars.  Near the maximum, there
are generically two temperatures, one below and one above the maximum,
that can fit the spectrum.  Autofit uses the photometry and detailed
line shape to choose between the two solutions, but it can happen that 
both have roughly equal likelihood.  In this case, two minima can be
seen on the likelihood contours on the summary plot.  Autofit reports
the likelihood weighted mean and second moment.  This means that it will
report a solution that is in between the two minima and not itself a good
fit.  However, the error will be large enough to include both minima.

At high temperatures ($>40,000$~K), our results will be biased
because we have used a model grid computed with LTE populations.  Stars
this hot are known to have significant non-LTE corrections that tend
to reduce the inferred temperature \citep{Nap99,Lie05}.  We have not
applied any corrections to our results.

Finally, we remind the reader that the autofit program was designed to
offer a first-pass estimate of temperatures and surface gravities and
to flag outliers.  It is no better than the details of fitting methodology
and the assumed atmosphere grid will allow.  While we are not aware of
any particular shortcomings save for those discussed above and in 
\S~\ref{sec:litcomp}, we caution
that the shear size of the sample does not insulate it from possible
small systematic biases in the recovered parameters.

\subsection{Rejected autofit values}

As noted above, we include a flag to signal that one should not use
the autofit values on some stars.
We include the autofit values on these stars {\it only} because
they might be helpful in finding stars that might have been
misclassified.  For example, a user interested in low gravity DAs
might wish to inspect the higher gravity subdwarfs to see if any
stars have been misclassified (in their opinion).  This would be
impossible if we omitted the autofit parameters from the hot subdwarf
list even though they are surely wrong for most of the stars therein.

When in doubt, we recommend that the reader check the figure showing
the model fit compared to the data.  If the fit looks bad, it probably is.

As noted above, the autofit temperatures and gravities are not correct
in detail for the hot subdwarfs because these stars often show mixed
element atmospheres, whereas we have only used single-element atmosphere
models.  However, it is likely that autofit does contain some information.
Figure \ref{fig:sd_ug_T} shows the distribution of autofit temperatures
and $u-g$ colors.  Many of the points fall along a clean locus.  Most of
the outliers are also outliers in the $u-g$ versus $g-r$ color plane.
This extra sequence is due to the presence of a main sequence 
companion that makes the $g-r$ color redder and also tends to dilute the
deep SD lines (B.~Green, private communication).  

\subsection{Testing autofit with duplicate spectra}

Because we have more than one spectrum of some of the white dwarfs, 
we have the opportunity to test the stability of the autofit results.
There are 775 duplicate spectra of white dwarfs in the catalog.
For 617 of these, the primary spectrum have the autofit quality flag
set, meaning that they are classified simply as DA or DB (possibly
with colons) and that the autofit model used appropriate element
in the atmosphere.  In 20 of these cases, the secondary spectrum fails
the autofit quality flag, and so we are left with 597 objects to compare.
These objects have completely separate spectra, taken on different nights,
often through different fibers on the spectrograph.  The autofit
analyses do share the same photometry.  Our general sense, however,
is that the photometry is very subdominant in the fitting and serves
primarily to break ties between multiple minima in the spectral fitting.

\begin{figure}[tb]
\plotone{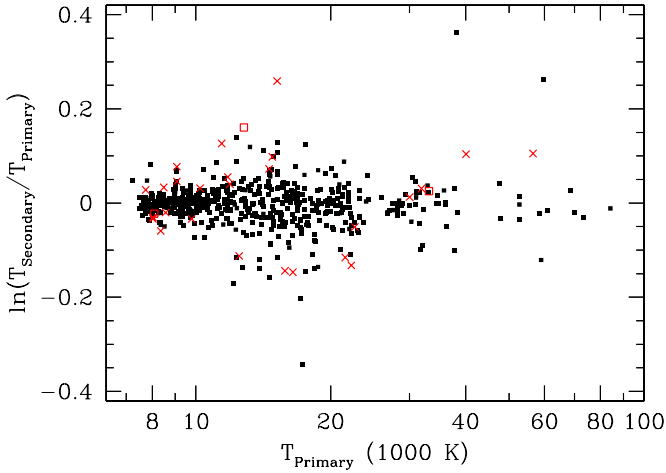}
\caption{\label{fig:dupT}%
For the duplicate spectra, the ratio in autofit
temperature between the two spectra (shown as natural
log of the ratio) versus the temperature of the primary
spectrum.  Only stars for which the autofit was considered
good in both observations are shown; this leaves 597 duplicate
observations.  Good means simply that the 
star was classified as a DA, DA:, DB, or DB: and that autofit
used the appropriate element for the model fit.  Stars with 3 observations
are shown twice, once for each secondary observation.  Objects whose
deviation is less than $3-\sigma$ for the quoted formal errors
are shown as black solid dots.  The 26 cases where the residuals
are between 3 and 5 $\sigma$ are shown as red crosses.  The 3 
cases where the residuals are more than 5 $\sigma$ are shown as
red open squares.  Note that most of the worst deviations are
properly reflected in the errors.
}
\end{figure}

Figure \ref{fig:dupT} shows the ratio in fitted temperatures as a
function of temperature.  In most cases, the agreement is close.  We
can also compute the formal error on the ratio.  This shows that in
most cases, the residuals are as expected from the autofit quoted errors.
Only 29 systems have more than 3 $\sigma$ residuals, and only 3 more than
5 $\sigma$.  Interestingly, the worst absolute residuals generally do
not have high statistical significance, meaning that the formal errors
are reporting the uncertainties.  Separating Figure \ref{fig:dupT}
into DA and DB stars doesn't change the behavior.

\begin{figure}[tb]
\plotone{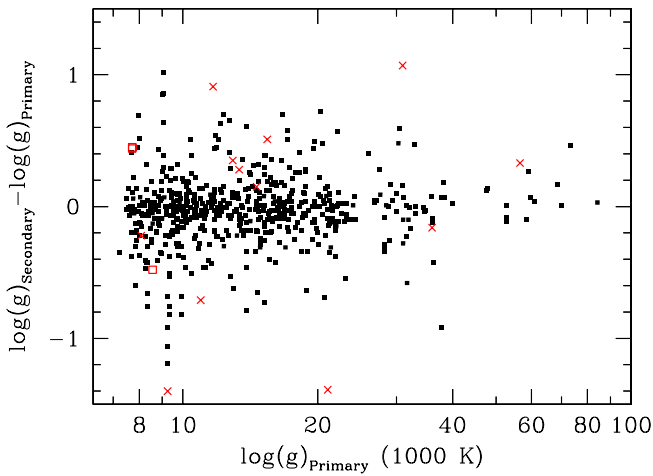}
\caption{\label{fig:dupG}%
As Figure \protect\ref{fig:dupT}, but for surface gravity.
The difference in $\log_{10} g$ is shown.  There are 3 objects
with residuals worse than 5 $\sigma$ and 12 between 3 and 5 $\sigma$.
Again, although some of the absolute residuals can be large,
these are typically reflected in the quoted errors.
}
\end{figure}

Figure \ref{fig:dupG} is a similar plot, but now for the difference in the
logarithm of the 
fitted surface gravity.  There are 15 3 $\sigma$ outliers and 3 5 $\sigma$
outliers.

\begin{figure}[tb]
\plotone{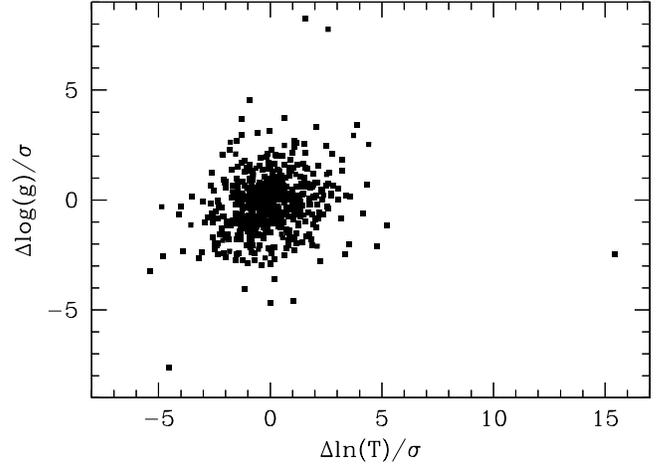}
\caption{\label{fig:dupTG}%
For the duplicate observations, the residuals in surface gravity
versus the residuals in temperature, both normalized by the formal
errors from autofit.  We use the natural logarithm of the ratio
of the two temperatures for the temperature residual, and the 
difference of $\log_{10} g$ for the surface gravity residual.
In both cases, the error $\sigma$ is that of the residual from
the combination of the formal errors.  All 597 duplicate observations
in which the autofit was considered good are shown.  Aside from
a few outliers, most observations lie in a tight locus that is 
close to a normal Gaussian.  In detail, the formal errors appear
to be underestimated by about 20\%.
}
\end{figure}

Finally, Figure \ref{fig:dupTG} shows the residuals in $\logg$ versus the
residuals in $T$, both normalized by the formal errors.  If the errors
were Gaussian, independent, and described by the formal errors, then this
would be a normal Gaussian distribution in both axes.  Aside from the handful
of outliers, this is not far from the case.  The central core appears to
be about 20\% broader than the unit normal, meaning that the formal errors
are about 20\% too small.  There is a hint that brighter stars have slightly
more scatter relative to their errors, say 40\% rather than 20\%, and all
5 of the 5 $\sigma$ outliers are brighter than $g=18$.  We repeat the caveat,
however,
about the errors being badly underestimated near the model grid boundaries.

It is worth noting that the secondary observations were not visually inspected.
The inspections of the primary observations were triggered by factors 
that could indicate data problems or suggest a classification other than
DA or DB.  If a data artifact was so bad as to prevent classification, then
the star would be excluded from this comparison.  However, aside from this,
data problems will remain in both the primary and secondary observation,
and only if the problems confused autofit into using the wrong element would
the star not enter this comparison.  Hence, we conclude that the formal
errors in the autofit method are properly incorporating 
the uncertainties from data problems.  This is in large part a
statement that the formal errors on the SDSS spectra are reasonable, at
least blueward of 7000\AA.  Of course, the results in this subsection on 
the stability of the autofit method only test the statistical errors and 
some observational systematics errors in the fitting.  Systematic errors from
the model atmospheres, from the autofit methodology, or from a generic bias in 
the SDSS dataset would not be revealed by comparing duplicate observations.

\subsection{Comparison to the Literature}
\label{sec:litcomp}

\begin{figure}[tb]
\plotone{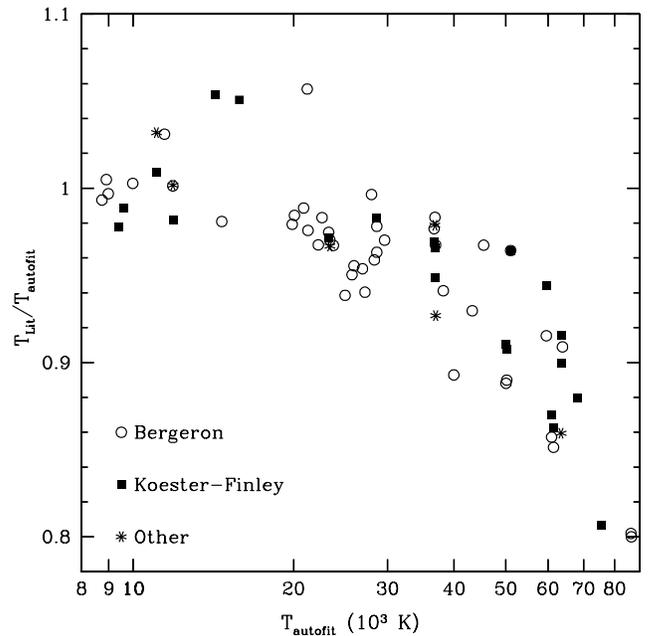}
\caption{\label{fig:litcomp_T}%
The comparison of autofit temperature to fits from the literature.
We see clearly that autofit temperatures become systematically higher
than the literature values for white dwarfs above 30,000~K.
The literature values have been separated into three groups to 
show the consistency in values between various groups.
Open circles are from
\citet{Ber92}, \citet{Ber94}, and \citet{Lie05}..
Filled circles are from \citet{Fin97}, \citet{Hom98}, and \citet{Koe01}.
Stars are from \citet{Mar97}, \citet{Ven97}, \citet{Nap99}.
}
\end{figure}

\begin{figure}[tb]
\plotone{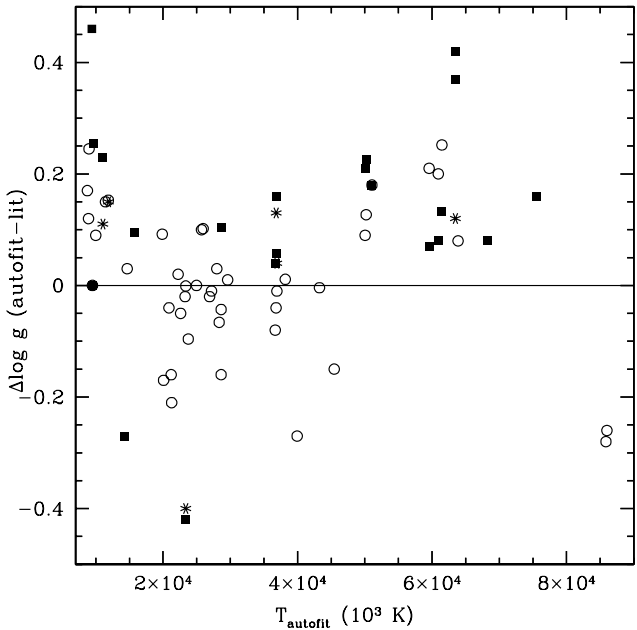}
\caption{\label{fig:litcomp_logg}%
The comparison of autofit surface gravities to fits from the literature,
as a function of temperature.
At low temperatures, autofit gravities are systematically high; this is
due to the modeling inconsistencies and grid artifacts discussed in
\S~\protect\ref{sec:autofitres}.
We also see a trend at high temperatures for autofit to be high.
This is related to the trend seem in the comparison of temperatures,
as there is a degeneracy in which displacements to higher temperatures
and higher gravities have compensating effects in the spectrum.
The symbols are the same as in Figure \protect\ref{fig:litcomp_T}.
}
\end{figure}

\begin{figure}[tb]
\plotone{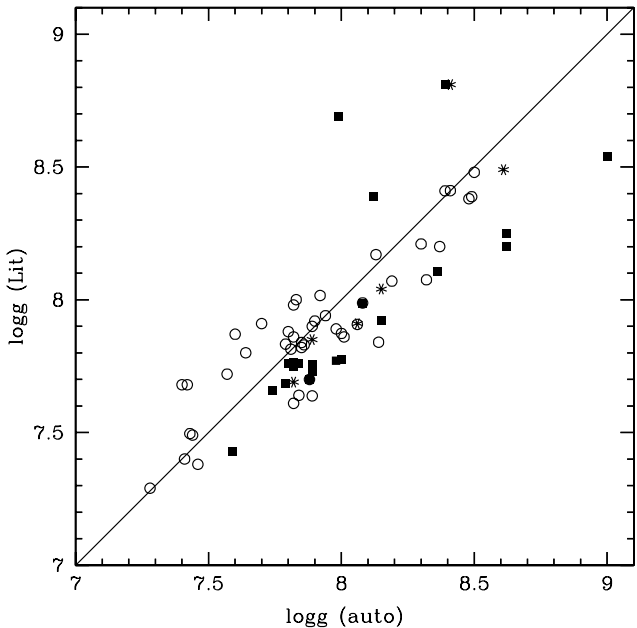}
\caption{\label{fig:litcomp_logg2}%
The comparison of autofit surface gravities to fits from the literature.
A high level of correlation is seen; in a broad sense, autofit is
recovering the same trends as other analyses.
The symbols are the same as in Figure \protect\ref{fig:litcomp_T}.
}
\end{figure}

We next compare our fits to those in the literature.  We find 71 fits
of 46 stars in the literature that we can compare to ours
\citep{Ber92,Ber94,Mar97,Ven97,Fin97,Hom98,Nap99,Koe01,Lie05}.
Figure \ref{fig:litcomp_logg2} compares the surface gravities
from the literature and autofit; the correlation implies that autofit
is recovering the same signal as other analyses.
Figures \ref{fig:litcomp_T} and \ref{fig:litcomp_logg}
show the residuals in the comparison as a function of temperature.
Below 30,000~K, the comparison suggests no systematic offset, save for
the fitting artifact in surface gravity for temperatures below 10,000~K.
However, for temperatures above 30,000~K, 
there is a clear systematic trend 
for autofit to be higher in temperature and in surface gravity
than the literature fits.  The residuals are at least 10\% in temperature
above 50,000~K.  We note that our modeling was already known to be wrong 
due to neglect of non-LTE physics and impure hydrogen atmospheres, but 
the literature fits are for the same assumptions and so should be comparable.

\begin{figure}[tb]
\plotone{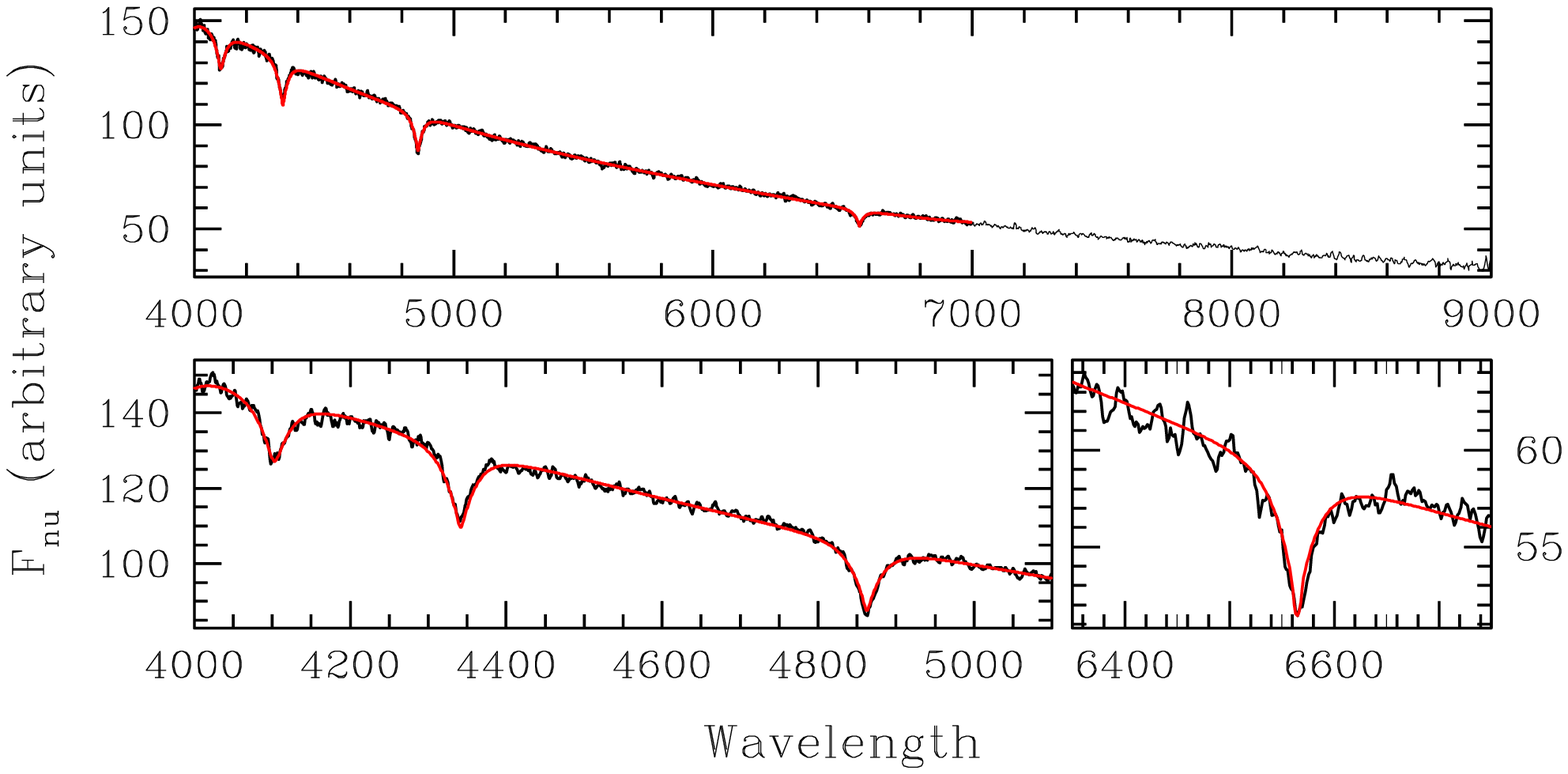} \\
\plotone{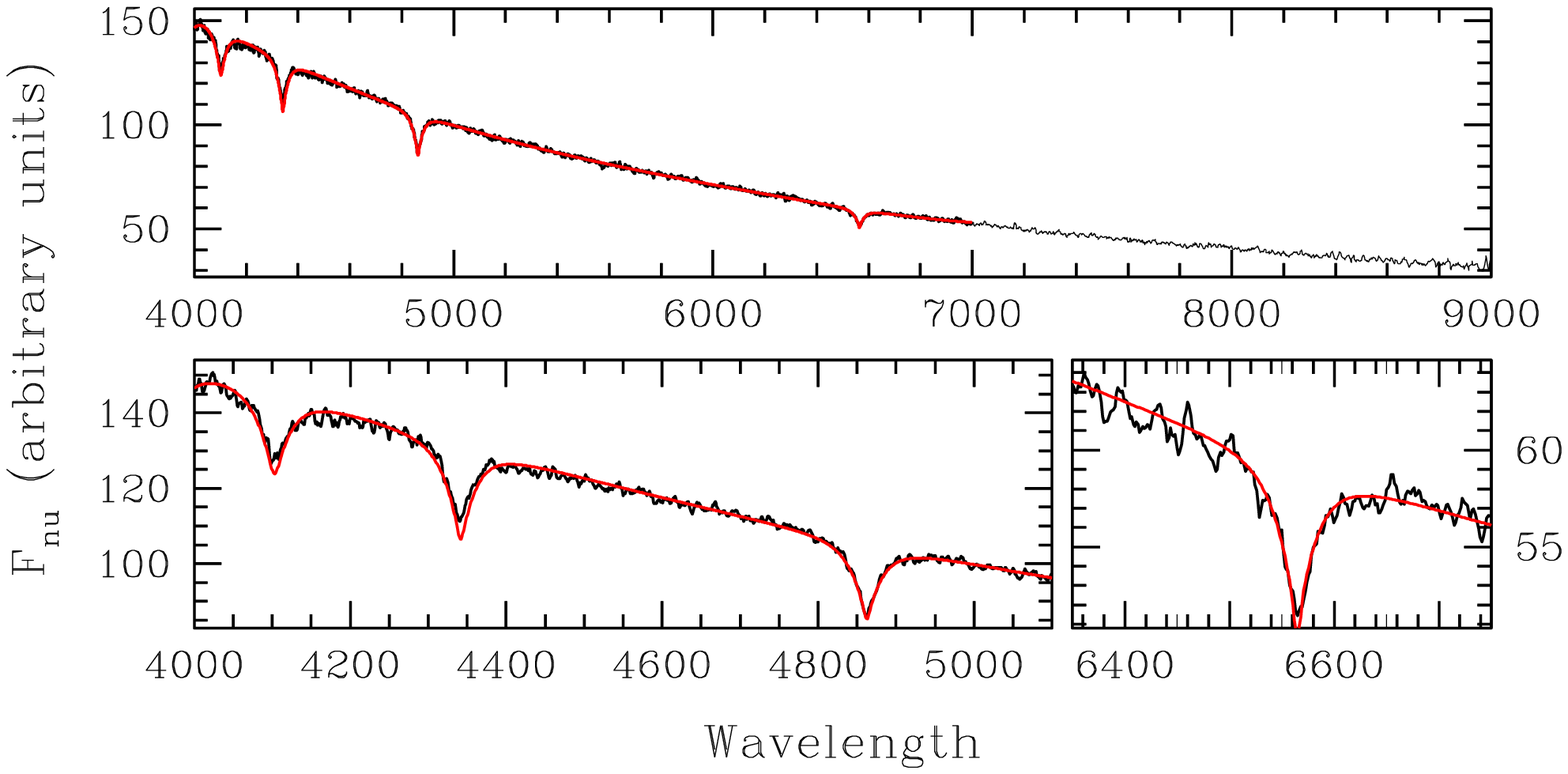}
\caption{\label{fig:wd1657}%
The spectrum and modeling for the hot DA SDSS J165851+341853, also
known as WD1657+343 \protect\citep{Gre86,McC99}, chosen to display the 
systematic shift between autofit and the literature for hot DA stars.  
The top two panels show
the spectrum overplotted with the best autofit fit of $T=60,943$~K,
$\logg=7.84$.  The bottom two panels show the spectrum overplotted
with $T=53,011$~K, $\logg=7.757$ \citep{Fin97}.  One sees that the
H$\gamma$ and H$\delta$ lines are too deep and too broad in the 
lower temperature model.  The higher temperature model is clearly 
the superior fit to the SDSS data given
our modeling assumptions and
there is no sign of any error in matching the continuum.
Indeed, the problem seems to be a small residual error in 
the SDSS spectrophotometry.
}
\end{figure}

We inspected our fits for these stars and found no sign of a systematic
problem in the autofit method.  For the SDSS data and the assumed models,
the fits appear reasonable.  Using the literature temperature and
gravities produce model H$\gamma$ and H$\delta$ lines that are clearly
deeper and slightly wider than the data.  There is no sign that the continuum fitting
is playing a role in the offset, and the autofit values do not change
substantially if we exclude data blueward of 4000\AA.  An example of 
this is shown for SDSS J165851+341853 (aka WD1657+343) in Figure 
\ref{fig:wd1657}.  

\begin{figure}[tb]
\plotone{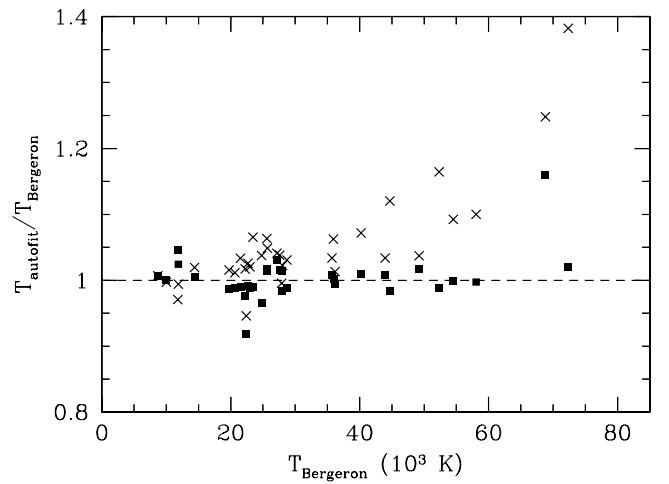}
\caption{\label{fig:Tcompare}
The comparison of autofit temperatures to fits from \citet{Lie05} for
two sets of spectra on a group of stars.  The crosses show the fits
to the SDSS spectra; the solid points show the fits to the \citet{Lie05}
spectra.  This demonstrates that autofit gives similar temperatures
to the fits of \citet{Lie05}, but that the two sets of spectra are
systematically different.  We believe that this is due to spectrophotometry
residuals in the SDSS at the H$\gamma$ and H$\delta$ lines.
}
\end{figure}

We then compared some of the spectra from \citet{Lie05} to the
corresponding SDSS spectra.  We found systematic offsets in the
depths of the H$\gamma$ and H$\delta$ lines, about 2\% relative to
the continuum, with a width of about 50\AA.  We believe that this
is due to small systematic errors in the SDSS spectrophotometry at
the locations of the Balmer lines, where the SDSS algorithm masks
the data when fitting the F subdwarf spectrophotometric stars.  When
we use the autofit program on the spectra from \citet{Lie05}, we
find temperatures and gravities that agree well with the literature.
The temperature comparison is shown in Figure \ref{fig:Tcompare};
the gravities do not shift and continue to agree.

These spectrophotometric residuals are small, but since they are
similar in breadth to the lines of a hot DA, they do affect the
temperatures.  The bias is quite small at lower temperatures when
the Balmer lines are strong, but becomes noticeable with the weaker
lines of the hottest DAs.  Of course, this effect is negligible 
for most non-WD applications; for narrow lines, most SDSS spectra
lack the signal-to-noise ratio to even detect a 2\% residual.

We are investigating methods to address this problem in future
catalogs.  At this point, users of our catalog should be cautious
about the exact temperature scale of the hot DA stars, even beyond
the usual caveats of non-LTE and multi-element atmosphere modeling.

\subsection{Completeness}
\label{sec:complete}

White dwarfs enter the SDSS spectroscopic sample through several different
targeting programs, all of which have different color, magnitude, and
photometric quality cuts as well as different priorities and sparse
samplings.  As such, we have not yet attempted to reconstruct in full
detail the completeness of the targeting.  However, we have done some
work on this topic, which we report in this section.  Note that this
section is only about what objects had a spectrum taken of them, not
whether we have properly identified the spectrum as that of a white dwarf.

We will assess the completeness by selecting from the DR4 imaging
catalogs all blue point sources with good photometry in all 5 bands
(good being defined by the ``clean sample'' set of flag checks 
defined in the DR4 on-line documentation).
After removing those objects that are not covered by a spectroscopic
plate, we then find the fraction of these targets of which a spectrum
was taken.  This fraction, as a function of magnitude, color, galactic
position, etc, is the completeness map.

We focus here on very blue point sources with colors typical of
single white dwarfs.  We restrict ourselves to $u-g<0.7$, $g-r<-0.1$,
and $r-i<-0.1$, dereddened.  This is where white dwarfs hotter than about 12,000~K
reside, but it is also the region where objects were excluded from
the optical quasar selection \citep{Ric02}, save for in the earliest
SDSS data.  
Objects with blue $u-g$ colors but redder colors in either $g-r$ or $r-i$
are targeted as quasars (down to $i<19.1$) with high completeness.  
Hence, the white dwarfs in this color region enter the 
sample by one of the stars, serendipity, or standards targeting packages.

An important thing to note is that all of the stars, serendipity, and
standards classes in the normal SDSS survey require that the target not
be blended on the sky with any other object.  By this, we do not mean
that the object must be successfully deblended (although most deblends are
successful), but that it could not have been blended in the first place,
meaning that the reference isophots cannot touch.  
The stars must also not be near the edge of a chip.
These are stringent cuts, intended to boost the purity of the sample.  
The exact flags are {\tt EDGE}, {\tt CHILD}, and {\tt BLENDED}; none
can be set.

We find that the fraction of sources with good photometry that pass the
not-blended cut is only about 60\%.  This fraction is mildly dependent
on galactic latitude: 55\% at $30^\circ$, 60\% at $40^\circ$, and 63\%
at $70^\circ$.  But it is strongly dependent on magnitude: 40\% at 
$g=15$, 55\% at $g=17$, 63\% at $18<g<19.5$.  The reason for this is
brighter stars reach the isophotal threshold at larger radius and hence
have more opportunity to be blended with fainter objects.  Closer to the
galactic plane, the latitude dependence does steepen as one would expect.

Focusing now on the non-blended stars, the completeness is still imperfect
because of the selection cuts and the sparse sampling.  The key set here
is the ``Hot\_Standard'' class, which requires $u-g<0$, $g-r<0$, and $g<19$,
dereddened \citep{Kle04}.
This class was observed with high priority as a tiled class \citep{Bla01t},
and the selection will include most of the brightest hottest stars.
However, the selection also required a fairly stringent photometric quality
flag, more stringent than the survey currently recommends.  As such, the
spectroscopic completeness for this color-magnitude box is 77\% of the 
non-blended stars.  For fainter stars in the range $19<g<19.5$, this
completeness drops to 50\%, as these stars enter from other stars and
serendipity selections.  We find no measurable dependence on $u-g$ in
either of these computations.

For stars at $0<u-g<0.7$, $g-r<-0.2$, $r-i<-0.1$, and $g<19.5$, dereddened, we find that 
the completeness is about 50\% of the non-blended stars, with no measurable
dependence on $u-g$.  For $-0.2<g-r<-0.1$, the completeness is mildly lower,
about 40\%.

The actual completenesses will be somewhat lower because our parent
sample required good photometry in all bands.  It is not correct to
compute this completeness correction by asking what fraction of bad
photometric objects got spectra, because many of the objects with bad
photometry are not hot stars but were simply scattered into the color
box as outliers.  Of course, bad things sometimes happen to good stars,
but one has to compute this fraction based on simulations of fake stars
through the photometric pipeline.  This failure branch is likely of order
5\%, but we have not quantified it in this paper. 

We also have not studied the completeness for $g>19.5$, for $u-g>0.7$, 
or for other red portions of color space, as this is not where the bulk
of our catalog lies.  

The SDSS does not target bright stars, as they saturate the spectrograph
and harm the extraction of traces from other fibers.  The stars and
serendipity packages all require the object to have fiber magnitudes
fainter than 15 in $g$, $r$, and $i$.  The PSF magnitudes are about 0.3
magnitudes brighter than the fiber magnitudes.  Completeness brighter
than $g<14.8$ will be poor, but we have not assessed this further.

Returning to the blended stars, the completeness within the SDSS is
below 10\%.  Blended stars can enter through the quasar or ROSAT target
algorithms.  Both have important color selections, and it is clear that
the blended stars in this WD catalog are typically much redder than the
unblended stars (closer to quasar colors).  With the completeness so low
and possessing an uncharacterized color bias, we recommend that blended
stars be avoided for statistical analyses.

We have marked the stars in the catalog as to whether they are unblended
or not.  Some of these blended stars have entered through the southern
special target selection, in which the quasar selection was performed
without the usual exclusion of the WD region \citep{Aba05}.  A few 
may have entered in the unusual case where the targeting was done on 
different imaging or earlier reductions than the ``best'' match in DR4
(which is what we use to report the blending status).

We note that binary stars typically do not get reported as blended.
Two point sources need to be a few arcseconds apart to be recognized
as two peaks by the deblender.  Many binary systems are unresolved in
the SDSS and simply appear as objects with unusual colors and spectra.
Intermediate systems, e.g., those 0.3-$3''$ apart, will be marked as
single objects but detected as extended.  These will then be rejected
by all stellar targeting classes, but picked up by galaxy targeting
down to $r=17.77$ \citep{Str02}.

In short, we expect that the completeness on hot single stars in the
range $14.8<g<19.5$ is essentially the product of a magnitude-dependent
factor of 40-60\% because of the not-blended criteria and another factor
of 40-77\% depending on color and magnitude.  Incompleteness of the
parent imaging catalog or of our spectroscopic classification are additional
factors, expected to be small for common classes of white dwarfs.

\section{Low-mass White Dwarf Candidates}
\label{sec:lowmass}

\citet{Lie04} presented two DA stars (and 6 other candidates)
with unusually low surface gravities and hence low masses.  SDSS
J123410.37-022802.9 in particular was found to be at 17,500~K and
$\logg=6.38\pm0.05$, implying a mass of $0.19\msun$.
Other examples of white dwarfs of similar mass
are MSP J1012+5307 \citep{Van96}, HD188112 \citep{Heb03}, and the
companion to PSF J0751+1807 \citep{Nic05}\footnote{This object is in
the SDSS DR4 area but is undetected to the limit of the SDSS imaging}.
White dwarfs with masses this small are thought to be composed of helium cores
rather than the canonical carbon-oxygen cores.  Such stars can be
formed when a companion strips the outer envelope from a helium-core
post-main-sequence star.  The existence of two of these rare stars around 
pulsars \citep{Van96,Nic05} suggests that the involvement of a neutron
star may be an important pathway for the formation of the low-mass white dwarf \citep{Lie04}.

Here we discuss 13 low-mass candidates from the DR4 catalog, 4 of which overlap
the list in \citet{Lie04}.  These objects are selected to have 
temperatures less than 30,000K, surface gravities more than
2 $\sigma$ below 7.2, and $g<20$.  We summarize the candidates in Table \ref{tab:lowmass}.

Most of the candidates are similar to SDSS J123410.37-022802.9 in
that they have temperatures between 15,000 and 20,000 K and gravities
between 6.2 and 7.0.  However, 3 of the candidates are significantly
lower temperature, suggesting even smaller masses.

\begin{figure}[tb]
\plotone{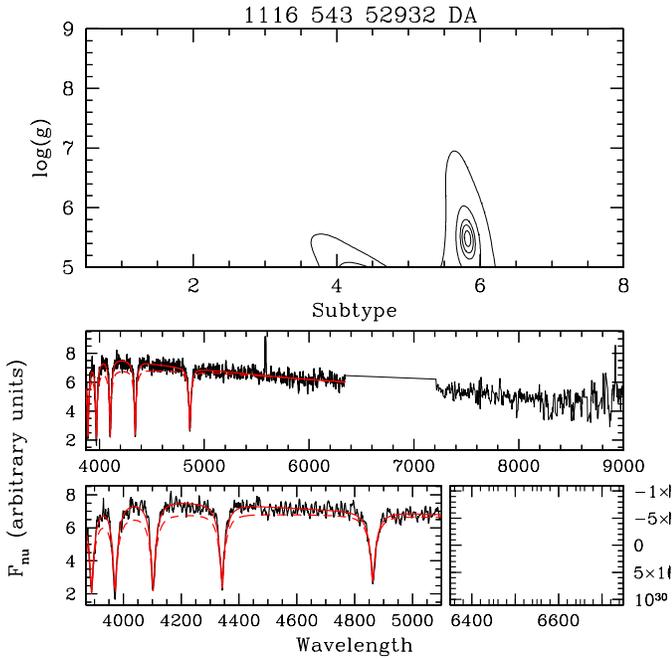}
\caption{\label{fig:lowmass1}%
The spectrum and autofit model for low-mass candidate SDSS J204949.78+000547.3.
The likelihood contours show evidence for multiple minima.  The best fit
to the spectroscopy is at $T=8660$~K and $\logg=5.48$, but the photometry
supports a higher temperature and lower gravity, closer to the secondary
minimum.
}
\end{figure}

SDSS J204949.78+000547.3 is our lowest gravity candidate.  The 
best fit to the spectroscopy is 8660 K with $\logg=5.48\pm0.10$.
However, as Figure \ref{fig:lowmass1} shows, the likelihood 
surface has a second minimum at roughly 11,500 K and even lower
gravity.  The primary photometry of this has a corrupted $g$ band,
which was excluded from the fit.  However, this star, being on
the equator in the south Galactic cap, has been imaged eight times
during the SDSS.  The averaged photometry (excluding the one
bad $g$ observation) is $u=20.29\pm0.02$, $g=19.20\pm0.01$,
$r=19.36\pm0.01$, $i=19.51\pm0.01$, and $z=19.60\pm0.04$.
The field is highly reddened, with a $g$-band extinction of 0.39 mag.
If we apply the full dereddening, then the colors become 
$u-g=-0.95$, $g-r=-0.27$, $r-i=-0.22$, and $i-z=-0.15$.  These
colors do not match the 8660~K, $\logg=5.48$ solution, and 
instead prefer the hotter, lower gravity solution.  In short, the
colors are more similar to a horizontal branch star, but the spectral lines are
clearly broader than that.  SDSS J204949.78+000547.3 remains 
unexplained.

\begin{figure}[tb]
\plotone{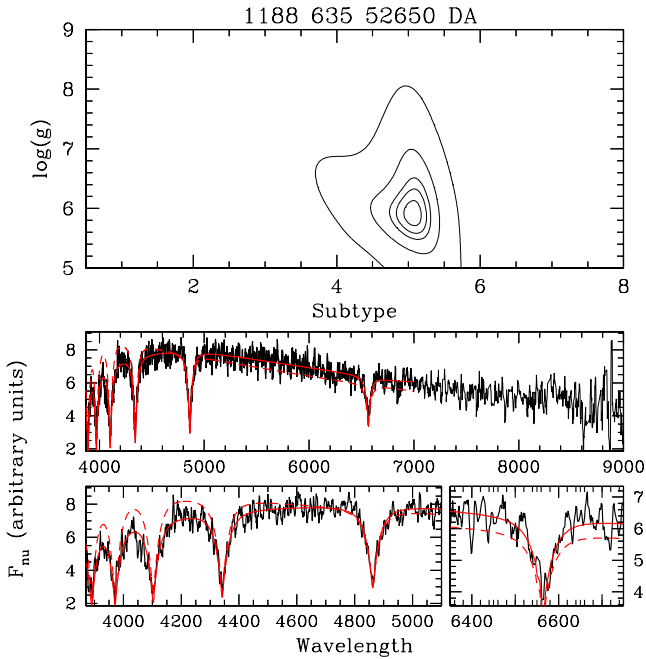}
\caption{\label{fig:lowmass2}%
The spectrum and autofit model for low-mass candidate SDSS J084910.13+044528.7.
The best fit to the spectrum is $T=9962$~K and $\logg=5.93$; the photometry
is consistent with these values.
}
\end{figure}

SDSS J084910.13+044528.7 (Fig.\ \ref{fig:lowmass2}) is a less extreme but more secure 
candidate.  The best fit is $9960\pm170$~K, $\logg=5.93\pm0.15$.
The dereddened colors of $u-g=0.88\pm0.05$, $g-r=-0.17\pm0.03$,
$r-i=-0.18\pm0.03$, $i-z=-0.29\pm0.08$ are a excellent match to 
the predicted colors of this fit.  The cooler temperature and
lower gravity than SDSS J123410.37-022802.9 
favor an even lower mass.  Of course, at 10,000~K, one is beginning
to reach the point where autofit gravities tend to be overestimated.
It is unknown whether the bias at higher gravities extends to 
lower gravities, but if it does, then the true gravity might be
somewhat lower than 5.93.  In any case, the spectrum of SDSS J084910.13+044528.7 
clearly has broader lines than a main sequence A star or blue horizontal
branch star.

Low temperature, low-mass stars such as these are incomplete in our
catalog simply because of the initial color cut, which was intended to
exclude main sequence and blue horizontal branch stars.  However, stars
of these colors do receive a healthy allocation of spectroscopic fibers
in the SDSS as part of the blue horizontal branch target selection.
We plan a more focused search for these stars in the future.

Two other candidates have multiple epochs of photometry.
SDSS J225242.25-005626.6 has been observed 9 times with photometry
$u=18.81\pm0.01$, $g=18.61\pm0.01$,
$r=18.85\pm0.01$, $i=19.07\pm0.01$, and $z=19.27\pm0.03$.
SDSS J002228.45+003115.5 has been observed 10 times with photometry
$u=19.51\pm0.03$, $g=19.34\pm0.02$,
$r=19.63\pm0.02$, $i=19.86\pm0.01$, and $z=20.11\pm0.05$.
Using this photometry in the fitting doesn't change the best fits.
We have included this photometry in Table \ref{tab:lowmass} but
not in the primary catalogs.

\begin{figure}[tb]
\plotone{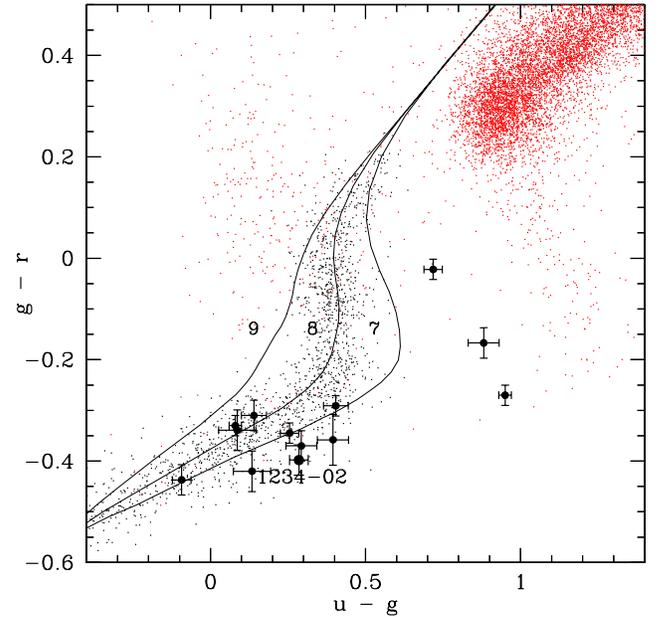}
\caption{\label{fig:lowmass_ugr}%
The $g-r$ and $u-g$ colors (dereddened) of the 13 low-mass candidates, overlaid on 
a background of DAs (small black points) as well as model predictions for DA colors at varying 
temperature at $\logg=7$, 8, and 9.  Small red points show a very sparse
sampling of main sequence stars, horizontal branch stars, and quasars
from \citet{Har01}.
See Figure \protect\ref{fig:ugrall} for the realistic distribution.
The confirmed object SDSS 1234--02 is labeled and has a somewhat larger
symbol.
Most of the candidates lie in the expected region of color space, but
three have more normal colors such that one would have to appeal to 
some mild photometry errors.  The low temperature cases lie well off the
DA locus but still somewhat blueward of the horizontal branch.
}
\end{figure}

\begin{figure}[tb]
\plotone{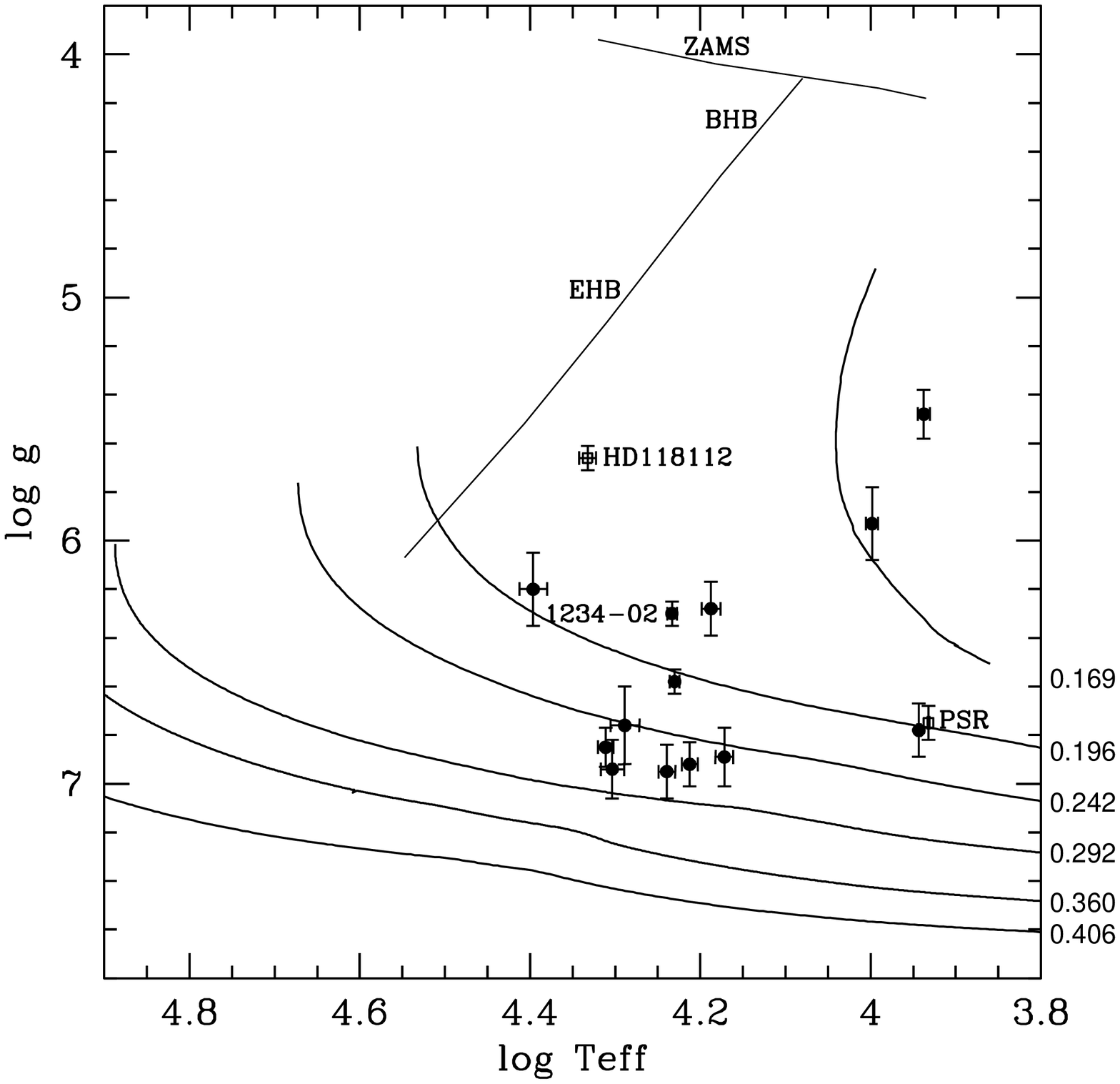}
\caption{\label{fig:lowmass_mass}%
The fitted surface gravity and temperatures of the 13 low-mass candidates, 
overlaid on tracks of constant mass from \protect\citet{Alt01}.
Also shown are the tracks of the zero-age main sequence (ZAMS) and horizontal branches (BHB and
EHB).
The confirmed objects 
MSP J1012+5307 \citep{Van96} 
HD188112 \citep{Heb03},
and SDSS 1234--02 \citep{Lie04} are labeled, the first as PSR because it is
around a known pulsar.
The low-temperature candidates may be the lowest mass white dwarfs yet found.
}
\end{figure}

Figure \ref{fig:lowmass_ugr} shows the $u-g$ vs. $g-r$ colors of these
stars, overlaid with the constant surface gravity model tracks.  
One sees that most of the
candidates do lie redward of the $\logg=7$ track in $u-g$, although in
three cases, one would need to invoke some mild photometric error.
Figure \ref{fig:lowmass_mass} shows the temperature and surface
gravity distribution overlaid with constant-mass predictions from
\citet{Alt01}.  These tracks suggest that these white dwarfs are below
$0.3\msun$, with 7 at or below $0.2\msun$.  The two stars discussed above,
SDSS J204949.78+000547.3 and SDSS J084910.13+044528.7,
may be the lowest mass white dwarfs yet found.
Given the recent discovery of a second massive pulsar around a helium-core
white dwarf \citep{Nic05}, these low-mass stars should be investigated
for signs of a neutron star.

In round numbers, models for these stars typically predict absolute magnitudes
of $M_g\sim8.5$.  This means that SDSS can recover the stars at
distances between 200 pc and 2 kpc.  At the high Galactic latitudes
of this catalog, we are probing out up to several disk scale lengths.
A detailed analysis of the space density of these stars will have to
wait for a full completeness analysis, but clearly the stars are rare,
with only 13 observed (times a yet-to-be-determined completeness factor)
over a survey volume of 4~kpc$^3$.

\section{The Hot Subdwarf Sequences}
\label{sec:sd}

The hot subdwarfs consist of two groups: 1) a sequence of generally helium-rich
sdO stars that have blackbody-like colors and largely coincide with
the DO-DB sequence, and 2) a sequence of hydrogen-rich sdB stars.
The latter are also called extended horizontal branch (EHB) stars when
they are found in globular and Galactic star clusters.

\begin{figure*}[p]
\centerline{\epsfysize=7in\epsffile{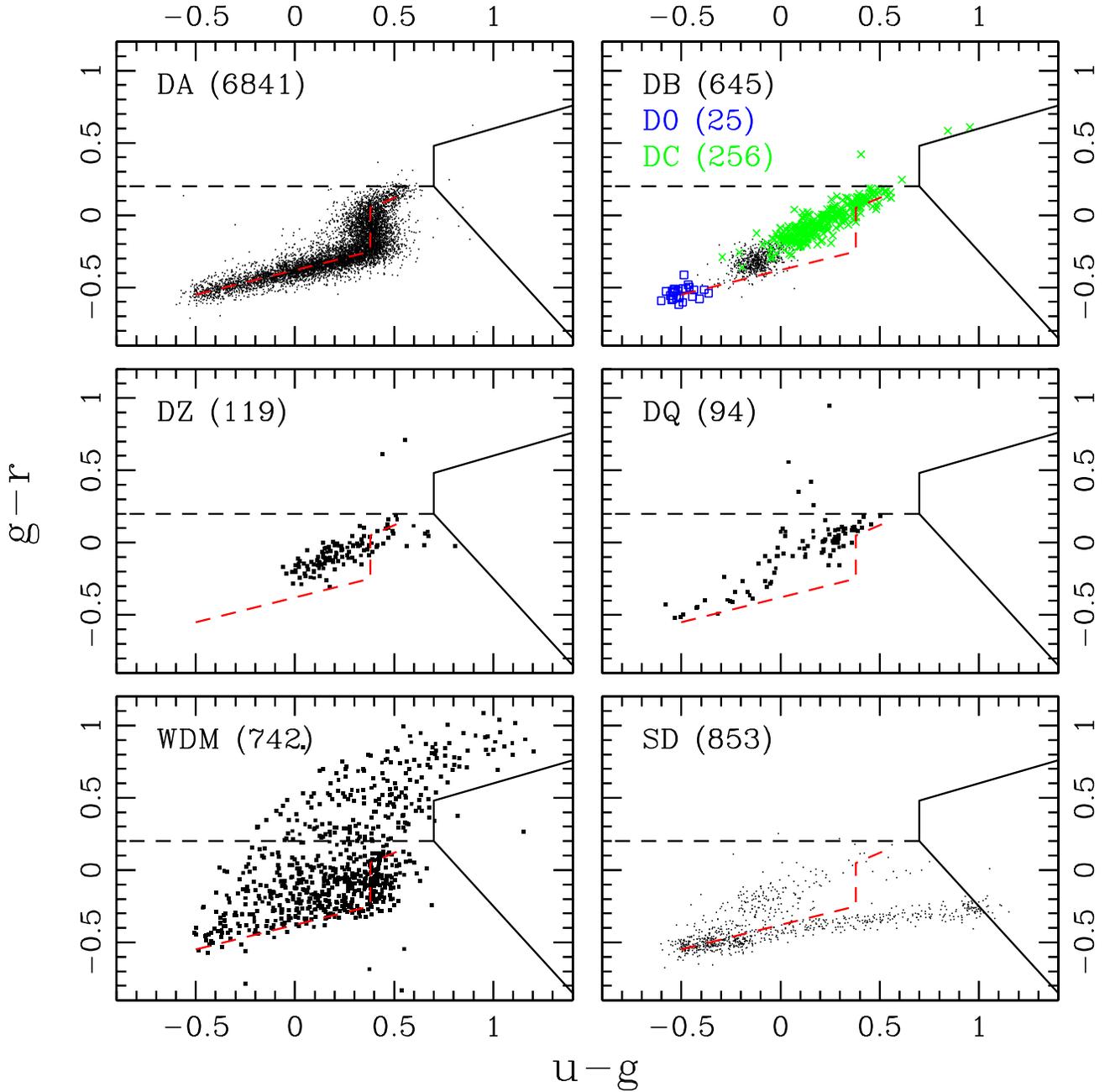}}
\caption{\label{fig:ugr}%
The $u-g$ vs $g-r$ colors of stars in sets according to their classification.
The photometry has been corrected for interstellar extinction \protect\citep{SFD}, assuming that the stars lie behind the full dust screen.  Stars that
are closer than that will be overcorrected and therefore be bluer on this 
plot than they intrinsically are.  Stars that are given an uncertain
classification are included in this plot.
The solid black lines shows the color cut applied to the selection of 
the sample.  The dashed line at $g-r=0.2$ is to remind the reader that
redward of this cut, the object is also required to lie off of the
main-sequence locus in $g-r$ and $r-i$.
{(\it upper left)}) DA classified stars.  An approximate fit to this locus
is reproduced on the other panels so that the reader may compare.
{(\it upper right)} DB, DO, and DC stars.
{(\it middle left)} DZ stars.
{(\it middle right)} DQ stars.
{(\it lower left)} All white dwarfs that are classified as having a companion.
As expected these stars lie redward than the single star locus, particularly
in $g-r$.  
{\it lower right)} Hot subdwarfs.  The locus tracking below the DA line 
is the single star locus; hot subdwarfs with companions form the sequence
that falls closer to the helium atmosphere line.  The single star locus
extends to rather red $u-g$ colors; however, our division between sdB
and BHB stars is probably too generous.  It is also inconsistent; 
there are many stars at $u-g\approx1$ that were rejected as A or BHB stars.
}
\end{figure*}

\begin{figure}[tb]
\plotone{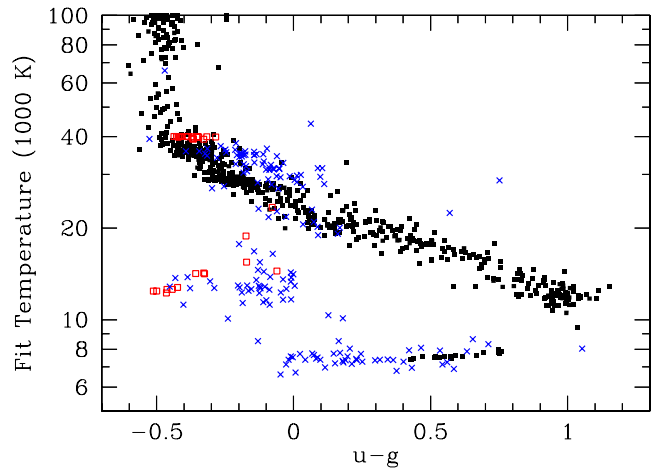}
\caption{\label{fig:sd_ug_T}%
The locus of $u-g$ color versus autofit temperature for stars classified
as hot subdwarfs.  The color is corrected for interstellar extinction assuming
that the star lies behind the \citet{SFD} prediction.  
Note that this is $u-g$ color, not $u-r$ as in 
Figures \protect\ref{fig:da_ur_T} and \protect\ref{fig:db_ur_T}.
We divide the single stars from those with companions by a cut of
$g-r<-0.3-0.2(u-g)$ (see Fig.~\protect\ref{fig:ugr}).
The black dots show single stars fit to hydrogen atmospheres.
These show a tight locus.  The temperature scale is surely wrong,
as hot subdwarfs commonly have mixed hydrogen-helium atmospheres
that would bias the results, but the autofitting is recovering 
a general temperature trend.  The red squares show single stars
fit to a helium atmosphere.  These temperatures have no correlation
with colors, and the set at 40,000K is due to the limits of the 
fitting.  Clearly, sdO atmospheres are not reliably matched by 
pure helium WD atmospheres.  The blue crosses show the locus of 
color versus temperature for the binary stars.  The 
spectroscopic fit obviously has enormous errors due to the change in line 
strengths for the composite spectra.
}
\end{figure}

The $u-g$ and $g-r$ colors of our hot subdwarfs are shown in the bottom,
right panel of Figure \ref{fig:ugr}.  Isolated stars run along the bottom
(bluer in $g-r$), with the sdO stars at the bluest tip.  The $u-g$ vs.\
$g-r$ color distribution shows two interesting features.  First, there
is a second locus in color that falls redward in $g-r$ of the primary
sequence.  These are subdwarfs with cool, nondegenerate binary companions.
These main-sequence stars redden the longer wavelength colors.  Unlike white dwarfs,
which are so faint as to be overwhelmed by anything but a M dwarf, the
subdwarfs can be blended with F or G main-sequence stars and still show
a very blue $u-g$ and $g-r$ color.  As shown in Figure \ref{fig:sd_ug_T},
the companion star does corrupt the spectral fit.

Second, the single star locus stretches continuously from the bluer ($u-g<0$)
end through to the edge of our selection at $u-g\approx1$ where the locus
joins that of the more common blue horizontal branch (BHB) and blue
straggler stars (BSS) near 10,000~K.  Traditionally, stars classified as
sdB have been hotter than 20,000~K; for example, the analyses of
\citet{Moe90} and \citet{Saf94} only included stars with derived $T_{\rm eff}\ge23,000$~K.
Such stars would likely be at $u-g<0$.
However, Figure \ref{fig:ugr} clearly shows that the hydrogen-rich
sequence extends redward and is continuous down to the BHB and BSS region,
such that the horizontal branch is populated from 10,000~K to near 40,000~K \citep{Gre74}.

There are two important complications in interpreting the distribution of stars
along this locus.
First, the SDSS target selection is not uniform across this region.
The ``Hot\_Standard'' class gives high priority to targets with $u-g<0$
and $g<19$.  Redder stars are still targeted, largely through the
blue-horizontal-branch selection \citep{Yan00,New02}, but the completeness
is not as high.  Second, the depth of the SDSS sample is such that 
different Galactic populations are being probed by the different types
of stars.  Notably, the BHB stars are a halo population while the 
EHB sample is primarily a disk population.  We are postponing
analysis of this locus to future work.

\section{Conclusions}

We have identified 9316 spectroscopically confirmed white dwarfs and
928 hot subdwarfs from the 800,000 spectra of the SDSS Data Release 4.
Rolling together all secondary classifications, the sample includes 8000 DA, 
713 DB, 289 DC, 41 DO/PG1159, 104 DQ, and 133 DZ stars.  There are of course
many different mixed and subclassifications.  More than 6000 of these
WDs are new, so this catalog roughly
doubles the number of spectroscopically confirmed white dwarfs.

We have fit model atmospheres to the DA and DB stars using our autofit
method.  We find the expected strong peak in surface gravity, along
with interesting outliers, such as the low-mass candidates discussed
in \S~\ref{sec:lowmass}.  The fitting results are validated for internal
consistency using
repeat observations and by correlations with the photometry.  There is
a systematic bias toward higher gravities at lower temperatures of
which users should beware.  This appears to be an artifact of the model
atmospheres.  There is also a bias toward over-estimating the temperature
of DA stars above 30,000~K.  This is caused a subtle systematic error
in the SDSS spectrophotometry causing the broad Balmer lines in the blue
to be slightly filled in.

Among our fitted stars, we continue to find a rare population of
low surface-gravity candidates that are very likely helium-core,
low-mass DA white dwarfs, building on the sample of \citet{Lie04}.
Two of these candidates may be the lowest mass DA stars yet found,
although this awaits confirmation.  These stars are very similar to
the two companions of massive pulsars \citep{Van96,Nic05}, opening the
hypothesis of a generic evolutionary role for neutron star companions
in the formation of these white dwarfs.

We cannot claim that this catalog is complete, particularly for difficult
cases such as DC stars, low-temperature DA stars, binary systems, and DH
stars.  For mundane warm to hot isolated DA, DB, and DO stars (above $\sim\!8,000~K$
for DA stars), 
we believe that the catalog should be reasonably complete (95\% or more)
within the SDSS spectroscopic catalog and down to spectral signal-to-noise
ratios where one can expect to classify the stars.  

In addition to searching for unusual classes of white dwarfs, large catalogs
open the possibility of large-number statistical studies of the distribution
of white dwarf properties as well as their Galactic structure.  Although
the SDSS spectroscopic target selection is not complete for these stars,
the incompletenesses can be modeled.  We have discussed the key aspects
of this modeling in \S~\ref{sec:complete}.

The SDSS DR4 represents roughly half of the final spectroscopic sample
for the original SDSS survey.  In addition, the new Sloan Extension
for Galactic Understanding and Exploration (SEGUE) is conducting an
extensive study at lower Galactic latitudes with both imaging and
spectroscopy.  Although white dwarfs are rare within the SDSS catalogs,
the available numbers are sufficient to open a key new opportunity to
study this diverse category of stars.

\bigskip

We thank Detlev Koester for providing grids of his recent model atmospheres.
We thank Betsy Green and Gary Schmidt for useful discussions.  We thank David Johnston
and Ryan Scranton for use of their stacked catalogs from the SDSS
equatorial region and Jay Holberg for help matching to the \citet{McC99}
catalog.
DJE was supported by NSF AST-0098577 and AST-0407200 and by an 
Alfred P.\ Sloan Research Fellowship.  JL was supported by NSF AST-0307321.

Funding for the Sloan Digital Sky Survey (SDSS) has been provided by the
Alfred P. Sloan Foundation, the Participating Institutions, the National
Aeronautics and Space Administration, the National Science Foundation,
the U.S. Department of Energy, the Japanese Monbukagakusho, and the Max
Planck Society. The SDSS Web site is http://www.sdss.org/.

The SDSS is managed by the Astrophysical Research Consortium (ARC)
for the Participating Institutions. The Participating Institutions are
The University of Chicago, Fermilab, the Institute for Advanced Study,
the Japan Participation Group, The Johns Hopkins University, the Korean
Scientist Group, Los Alamos National Laboratory, the Max-Planck-Institute
for Astronomy (MPIA), the Max-Planck-Institute for Astrophysics (MPA),
New Mexico State University, University of Pittsburgh, University of
Portsmouth, Princeton University, the United States Naval Observatory,
and the University of Washington.

{}

\clearpage

\clearpage

\clearpage

\clearpage

\clearpage

\clearpage

\begin{table}[tb]\footnotesize
\caption{\label{tab:wdsuper}}
\begin{center}
{\sc White Dwarf Dominant Classifications\\}
\begin{tabular}{lr} 
\tableskip\tableline\tableline\tableskip
Classification$^{a}$ & Number \\
\tableskip\tableline\tableline\tableskip
DA & 8000 \\
DB & 713 \\
DC & 289 \\
DH & 9 \\
DO & 31 \\
DQ & 104 \\
DZ & 133 \\
PG1159 & 10 \\
WD & 27 \\

\tableskip\tableline\tableline\tableskip
\end{tabular}
\end{center}
NOTES.---%
$^{a}$ This classification refers to the dominant
classification, given by the first two letters of
the full classification.  WD means that the type
of the WD was not clear.
\end{table}

\begin{table}[tb]\footnotesize
\caption{\label{tab:wd}}
\begin{center}
{\sc White Dwarf Classifications\\}
\begin{tabular}{lrlr} 
\tableskip\tableline\tableline\tableskip
Class & Number & Class & Number \\
\tableskip\tableline\tableline\tableskip
DA & 2080 & DB+M: & 5 \\
DA: & 163 & DB O: & 1 \\
DA auto & 4919 & DB: or HeSDB & 3 \\
DAB & 7 & DB:+QSO: & 1 \\
DA B: & 2 & DBZ & 8 \\[3pt]
DAB: & 1 & DB Z: & 1 \\
DABH & 1 & DBZ: & 1 \\
DA BH: & 1 & DBZA & 3 \\
DA CV: & 4 & DBZ A: & 2 \\
DA+dC & 1 & DC & 142 \\[3pt]
DAE & 1 & DC: & 131 \\
DAH & 46 & DCA: & 1 \\
DA H: & 10 & DC:+M & 5 \\
DAH+DA: & 1 & DC+M & 8 \\
DA H:+M: & 1 & DC+M: & 2 \\[3pt]
DAH+M: & 1 & DH & 5 \\
DAHpec & 2 & DH: & 3 \\
DA+K: & 1 & DH+DA & 1 \\
DA:+M & 30 & DO & 19 \\
DA:+M: & 5 & DO: & 9 \\[3pt]
DA+M & 611 & DO+M & 3 \\
DA+M: & 81 & DQ & 45 \\
DAO & 13 & DQ: & 26 \\
DAOB & 2 & DQABhot & 1 \\
DApec & 3 & DQAhot & 3 \\[3pt]
DApec: & 2 & DQHhot & 1 \\
DA Q: & 1 & DQhot & 18 \\
DA+QSO: & 1 & DQhot: & 2 \\
DA:+sdK: & 1 & DQhot H: & 2 \\
DA+sdK & 1 & DQ:+M: & 1 \\[3pt]
DAZ & 4 & DQpec & 5 \\
DAZ: & 3 & DZ & 95 \\
DB & 508 & DZ: & 17 \\
DB: & 54 & DZA & 13 \\
DBA & 56 & DZ A: & 1 \\[3pt]
DB A: & 28 & DZA+M & 1 \\
DBA: & 12 & DZB & 1 \\
DBA:+M & 1 & DZBA & 2 \\
DBA+M & 1 & DZ:+M & 1 \\
DBApec: & 1 & DZ+M & 2 \\[3pt]
DBAZ & 6 & PG1159 & 10 \\
DB balmerE & 1 & WD & 1 \\
DBH & 3 & WD:+M & 22 \\
DB:+M & 2 & WD+M & 2 \\
DB+M & 15 & WD+M: & 2 \\[3pt]

\tableskip\tableline\tableline\tableskip
\end{tabular}
\end{center}
\end{table}

\begin{table}[tb]\footnotesize
\caption{\label{tab:sd}}
\begin{center}
{\sc Hot Subdwarf Classifications\\}
\begin{tabular}{lr} 
\tableskip\tableline\tableline\tableskip
Classification & Number \\
\tableskip\tableline\tableline\tableskip
HeSDB & 5 \\
HeSDB: & 11 \\
SD: & 8 \\
SD auto & 391 \\
SDB & 195 \\
SDB: & 31 \\
SDB+G & 4 \\
SDB+M & 3 \\
SDB+M: & 1 \\
SDB+MS: & 7 \\
SD:+M & 1 \\
SDO & 218 \\
SDO: & 52 \\
SDO+G: & 1 \\

\tableskip\tableline\tableline\tableskip
\end{tabular}
\end{center}
\end{table}

\begin{table*}[tb]\footnotesize
\caption{\label{tab:cat}}
\begin{center}
{\sc Format of Catalog Table\\}
\begin{tabular}{rl}
\tableskip\tableline\tableline\tableskip
Column & Description \\
\tableskip\tableline\tableline\tableskip
1& Classification \\
2& \protect\citet{McC99} WD Catalog Identification Label (as of August 2005) \\
3& Provenance of the SDSS classification$^a$ \\
4& Primary spectrum, $=0$ if a duplicate spectrum, $=1$ otherwise \\
5& SDSS IAU-style name$^b$ \\
6& Right Ascension (J2000)$^c$ \\
7& Declination (J2000)$^c$ \\
8& Proper Motion (arcsec/century) \\
9& Proper Motion position angle (degrees), $=0$ for north, $=90$ for east \\
10--11& $\Delta$RA, $\Delta$Dec Proper Motion (arcsec/century) \\
12& Time of SDSS imaging observation (MJD)$^c$ \\
13& SDSS Run number \\
14& SDSS Rerun number \\
15& SDSS Camera Column \\
16& SDSS Field number \\
17& SDSS ID number$^{d}$ \\
18--22& $u$, $g$, $r$, $i$, $z$ magnitudes (PSF, as observed) \\
23--27& $u$, $g$, $r$, $i$, $z$ magnitude errors  \\
28--32& $u$, $g$, $r$, $i$, $z$ photometry flag summary ($=1$ for bad, $=0$ for good) \\
33& Extinction in the $g$ band$^e$ \\
34& Plate number \\
35& Fiber number \\
36& MJD number for spectroscopic observation \\
37& $S/N$ of spectrum in the $g$ band, per spectroscopic pixel \\
38& Deblended?  $=1$ if the object was isolated, $=0$ if it was deblended.$^f$ \\
39& PrimTarget flag (decimal)$^g$ \\
40& PrimTarget flag (hex)$^g$ \\
41& SecTarget flag (decimal)$^g$\\
42& SecTarget flag (hex)$^g$\\
43& Autofit quality flag.  $=1$ if good.  
If zero, do not use the autofit numbers for anything but diagnostic searching \\
44& Autofit atmosphere element ($=1$ for hydrogen, $=2$ for helium) \\
45& Autofit Temperature (Kelvin) \\
46& Autofit Temperature Error (Kelvin) \\
47& Autofit \logg\ (dex) \\
48& Autofit \logg\ Error \\
49& Autofit $\chi^2$ per degree of freedom \\
\tableskip\tableline\tableline\tableskip
\end{tabular}
\end{center}
NOTES.---%
$^a$ This is merely to track the provenance of our spectral classification;
it does not indicate discovery nor literature classifications, which could
be tracked through the \citet{McC99} catalog.
Kle04: \protect\citet{Kle04}
Sil05: \protect\citet{Sil05}
DR4: This paper. \\
$^b$ This is the name from the SDSS DR4.  Because of tiny astrometry shifts, it is 
possible that this name may differ from the name in previous SDSS releases.  This 
breaks the IAU convention that published names never change.
We strongly recommend that associations be performed on the astrometric coordinates
rather than on the name. \\
$^c$ The epoch of the coordinates is given by the time of the SDSS imaging
observation in column 12. \\
$^d$ ID number within the field, not the objID in the SDSS CAS.\\
$^e$ The canonical SDSS extinction curve in $u$, $g$, $r$, $i$, and $z$ is 1.36, 1.00, 0.73, 0.55, and 0.39 times $A_g$. \\
$^f$ Most of the relevant target selection catagories require that
the object be sufficiently isolated on the sky that it was not blended with
any other object.  Blended stars have much lower completeness; see 
\S~\protect\ref{sec:complete}. \\
$^g$ These bit fields record the decisions from the various spectroscopic targeting algorithms.  See \protect\citet{Sto02} and \protect\citet{Aba05} or the on-line documentation
for details.
The southern survey bit (0x80000000) has been suppressed in the secTarget flag
for brevity but retained in the primTarget flag \protect\citep{Aba05}.
\end{table*}

\begin{table*}[tb]\footnotesize
\caption{\label{tab:dr1orphans}}
\begin{center}
{\sc List of Stars from DR1 Catalog Missing from this Catalog\\}
\begin{tabular}{llllrrl}
\tableskip\tableline\tableline\tableskip
Name & Plate & MJD & Fiber & RA (J2000) & Decl (J2000) & Class \\
\tableskip\tableline\tableline\tableskip
SDSS J020848.28+121332.4 & 0428 & 51883 & 246 &  32.20118 &  12.22568 & DA \\
SDSS J144738.39+034930.4 & 0587 & 52026 & 118 & 221.90997 &   3.82511 & DA \\
SDSS J102448.85--002312.3 & 0272 & 51941 & 180 & 156.20354 &  --0.38675 & DA: \\
SDSS J124337.50+671252.1 & 0494 & 51915 & 033 & 190.90623 &  67.21448 & DA: \\
SDSS J170919.90+612016.8 & 0351 & 51780 & 112 & 257.33293 &  61.33799 & DA: \\
SDSS J102414.85+655551.6 & 0489 & 51930 & 338 & 156.06186 &  65.93099 & DA6 \\
SDSS J080755.99+485419.3 & 0440 & 51885 & 275 & 121.98330 &  48.90537 & DA7 \\
SDSS J104912.65--000854.8 & 0275 & 51910 & 110 & 162.30270 &  --0.14855 & DA7 \\
SDSS J115052.32+683116.1 & 0492 & 51955 & 523 & 177.71800 &  68.52113 & DA7 \\
SDSS J171713.11+574634.6 & 0355 & 51788 & 548 & 259.30463 &  57.77628 & DA7 \\
SDSS J105628.49+652313.5 & 0490 & 51929 & 205 & 164.11870 &  65.38707 & DAH \\
SDSS J010352.23--100230.2 & 0659 & 52199 & 318 &  15.96764 & --10.04172 & DA:H: \\
SDSS J001733.59+004030.4 & 0389 & 51795 & 614 &   4.38997 &   0.67512 & DAM \\
SDSS J032428.78--004613.8 & 0414 & 51901 & 263 &  51.11991 &  --0.77050 & DAM \\
SDSS J081410.59+452315.7 & 0439 & 51877 & 583 & 123.54414 &  45.38770 & DAM \\
SDSS J172043.87+560109.3 & 0367 & 51997 & 464 & 260.18281 &  56.01925 & DAM \\
SDSS J155340.36+011335.3 & 0343 & 51692 & 367 & 238.41816 &   1.22647 & DB \\
SDSS J160437.77+005206.7 & 0344 & 51693 & 523 & 241.15736 &   0.86853 & DB \\
SDSS J091436.46+014514.5 & 0473 & 51929 & 319 & 138.65192 &   1.75403 & DB: \\
SDSS J154938.64--001318.2 & 0342 & 51691 & 104 & 237.41101 &  --0.22173 & DB: \\
SDSS J173327.35+585439.8 & 0366 & 52017 & 582 & 263.36394 &  58.91106 & DB\_UNC \\
SDSS J000116.53+000205.4 & 0387 & 51791 & 177 &   0.31887 &   0.03483 & DC \\
SDSS J000918.12--002141.5 & 0388 & 51793 & 145 &   2.32550 &  --0.36152 & DC \\
SDSS J005253.74+135618.2 & 0420 & 51871 & 207 &  13.22390 &  13.93839 & DC \\
SDSS J012048.78--090740.9 & 0661 & 52163 & 393 &  20.20323 &  --9.12802 & DC \\
SDSS J012723.62--004630.0 & 0399 & 51817 & 099 &  21.84843 &  --0.77500 & DC \\
SDSS J025457.18--072641.8 & 0457 & 51901 & 483 &  43.73826 &  --7.44494 & DC \\
SDSS J040409.73--044909.1 & 0465 & 51910 & 359 &  61.04056 &  --4.81918 & DC \\
SDSS J073515.12+362445.0 & 0431 & 51877 & 007 & 113.81302 &  36.41249 & DC \\
SDSS J080623.77+430519.9 & 0437 & 51869 & 547 & 121.59905 &  43.08887 & DC \\
SDSS J082637.86+503357.3 & 0442 & 51882 & 142 & 126.65773 &  50.56593 & DC \\
SDSS J085430.40--001836.5 & 0468 & 51912 & 099 & 133.62665 &  --0.31014 & DC \\
SDSS J085826.04--001416.8 & 0469 & 51913 & 124 & 134.60849 &  --0.23801 & DC \\
SDSS J090725.13+581821.5 & 0484 & 51907 & 224 & 136.85471 &  58.30597 & DC \\
SDSS J091652.16+554338.9 & 0451 & 51908 & 241 & 139.21734 &  55.72748 & DC \\
SDSS J094400.65+562543.3 & 0556 & 51991 & 525 & 146.00270 &  56.42869 & DC \\
SDSS J094821.69+594933.9 & 0453 & 51915 & 538 & 147.09038 &  59.82609 & DC \\
SDSS J102506.14+650332.8 & 0489 & 51930 & 303 & 156.27557 &  65.05910 & DC \\
SDSS J102856.58+640700.1 & 0489 & 51930 & 289 & 157.23574 &  64.11669 & DC \\
SDSS J104523.87+015722.1 & 0506 & 52022 & 113 & 161.34944 &   1.95613 & DC \\
SDSS J113844.07+643848.9 & 0597 & 52059 & 559 & 174.68362 &  64.64692 & DC \\
SDSS J133739.40+000142.9 & 0299 & 51671 & 357 & 204.41418 &   0.02857 & DC \\
SDSS J134419.47+670255.8 & 0497 & 51989 & 448 & 206.08111 &  67.04884 & DC \\
SDSS J134821.45+653013.8 & 0497 & 51989 & 148 & 207.08937 &  65.50384 & DC \\
SDSS J150516.97+001659.5 & 0310 & 51990 & 517 & 226.32069 &   0.28320 & DC \\
SDSS J151136.74+015518.1 & 0540 & 51996 & 518 & 227.90310 &   1.92169 & DC \\
SDSS J160010.06+005513.1 & 0344 & 51693 & 323 & 240.04190 &   0.92031 & DC \\
SDSS J165401.26+625355.0 & 0349 & 51699 & 220 & 253.50523 &  62.89860 & DC \\
SDSS J172400.74+573538.2 & 0366 & 52017 & 248 & 261.00309 &  57.59394 & DC \\
SDSS J231604.93--005350.2 & 0382 & 51816 & 215 & 349.02054 &  --0.89728 & DC \\
SDSS J031729.73--071644.7 & 0460 & 51924 & 319 &  49.37389 &  --7.27909 & DC: \\
SDSS J094525.58+585916.3 & 0453 & 51915 & 190 & 146.35658 &  58.98786 & DC: \\
SDSS J104528.96+654009.2 & 0489 & 51930 & 586 & 161.37065 &  65.66922 & DC: \\
SDSS J144240.73+610152.2 & 0608 & 52081 & 414 & 220.66971 &  61.03117 & DC: \\
SDSS J144816.80+003757.6 & 0537 & 52027 & 043 & 222.07000 &   0.63266 & DC: \\
SDSS J145357.70+605851.3 & 0611 & 52055 & 313 & 223.49040 &  60.98091 & DC: \\
SDSS J211841.07--080347.0 & 0639 & 52146 & 126 & 319.67111 &  --8.06305 & DC: \\
SDSS J142830.48+013443.6 & 0534 & 51997 & 022 & 217.12698 &   1.57879 & DCM \\
SDSS J035222.86--060506.3 & 0463 & 51908 & 559 &  58.09523 &  --6.08510 & DQ \\
SDSS J101219.90+004019.7 & 0502 & 51957 & 095 & 153.08290 &   0.67214 & DQ \\
SDSS J135628.25--000941.2 & 0301 & 51942 & 231 & 209.11770 &  --0.16144 & DQ \\
SDSS J205316.34--070204.3 & 0636 & 52176 & 267 & 313.31809 &  --7.03453 & DQ \\
SDSS J234132.83--010104.5 & 0385 & 51877 & 126 & 355.38679 &  --1.01793 & DQ \\
SDSS J025322.47--080858.2 & 0457 & 51901 & 170 &  43.34364 &  --8.14951 & DQ: \\
SDSS J133359.86+001654.8 & 0298 & 51955 & 492 & 203.49941 &   0.28189 & DQH \\
SDSS J121837.12+002304.0 & 0288 & 52000 & 423 & 184.65466 &   0.38443 & DZ \\
SDSS J140316.91--002450.0 & 0301 & 51942 & 030 & 210.82045 &  --0.41388 & DZMG \\
SDSS J155752.42+523514.4 & 0618 & 52049 & 007 & 239.46842 &  52.58732 & SDB \\
SDSS J204403.98--051135.6 & 0635 & 52145 & 360 & 311.01658 &  --5.19323 & SDB \\
SDSS J031303.31--061600.5 & 0459 & 51924 & 536 &  48.26378 &  --6.26681 & SDB: \\
SDSS J212053.73--071544.1 & 0640 & 52178 & 310 & 320.22387 &  --7.26226 & SDO \\
SDSS J040723.78--044718.8 & 0465 & 51910 & 432 &  61.84908 &  --4.78855 & WDDC: \\
SDSS J161044.47+004650.5 & 0345 & 51690 & 532 & 242.68528 &   0.78070 & WDDZ: \\

\tableskip\tableline\tableline\tableskip
\end{tabular}
\end{center}
\end{table*}

\begin{table*}[tb]\footnotesize
\caption{\label{tab:lowmass}}
\begin{center}
{\sc Low-mass White Dwarf Candidates\\}
\begin{tabular}{lrrccccccrcc}
\tableskip\tableline\tableline\tableskip
Name & Plate & Fiber & MJD & $g$ & $u-g$ & $g-r$ & $r-i$ & $i-z$ & $T$ (K) & $\logg$ & $\sigma(\logg)$ \\
\tableskip\tableline\tableline\tableskip
SDSS J204949.78+000547.3 & 1116 & 543 & 52932 & 18.81 & +0.95 & --0.27 & --0.22 & --0.15 &  8660 & 5.48 & 0.10 \\
SDSS J084910.13+044528.7 & 1188 & 635 & 52650 & 19.11 & +0.88 & --0.17 & --0.17 & --0.29 &  9962 & 5.93 & 0.15 \\
SDSS J162542.10+363219.1 & 1338 & 427 & 52765 & 19.31 & --0.09 & --0.44 & --0.34 & --0.17 & 24913 & 6.20 & 0.15 \\
SDSS J105353.89+520031.0 & 1010 &  12 & 52649 & 18.87 & +0.40 & --0.36 & --0.23 & --0.43 & 15399 & 6.28 & 0.11 \\
SDSS J123410.36--022802.8$^a$ &  335 & 264 & 52000 & 17.72 & +0.28 & --0.40 & --0.24 & --0.29 & 17114 & 6.30 & 0.05 \\
SDSS J143633.29+501026.8 & 1046 & 594 & 52460 & 18.16 & +0.26 & --0.34 & --0.29 & --0.32 & 16993 & 6.58 & 0.05 \\
SDSS J002207.65--101423.5$^a$ &  653 & 225 & 52145 & 19.62 & +0.09 & --0.34 & --0.23 & --0.33 & 19444 & 6.76 & 0.16 \\
SDSS J082212.57+275307.4 & 1267 & 352 & 52932 & 18.22 & +0.72 & --0.02 & --0.13 & --0.16 &  8777 & 6.78 & 0.11 \\
SDSS J225242.25--005626.6 &  676 & 138 & 52178 & 18.28 & +0.08 & --0.33 & --0.28 & --0.25 & 20479 & 6.85 & 0.08 \\
SDSS J163030.58+423305.7 &  815 &  36 & 52374 & 18.98 & +0.40 & --0.29 & --0.32 & --0.34 & 14854 & 6.89 & 0.12 \\
SDSS J142601.47+010000.2$^a$ &  305 & 512 & 51613 & 19.22 & +0.29 & --0.37 & --0.26 & --0.23 & 16311 & 6.92 & 0.09 \\
SDSS J105611.02+653631.5$^a$ &  490 & 215 & 51929 & 19.67 & +0.13 & --0.42 & --0.25 & --0.19 & 20112 & 6.94 & 0.12 \\
SDSS J002228.45+003115.5 &  688 & 466 & 52203 & 19.25 & +0.14 & --0.31 & --0.25 & --0.26 & 17355 & 6.95 & 0.11 \\

\tableskip\tableline\tableline\tableskip
\end{tabular}
\end{center}
NOTES.---%
$^a$ Also presented in \protect\citet{Lie04}. \\
All magnitudes and colors are dereddened.
\end{table*}

\end{document}